\documentclass{JHEP3}
\usepackage{epsfig,amsmath,bbm,subfigure}
\usepackage{dsfont}
\usepackage{wasysym}
\usepackage{rotating}
\usepackage{cite}

\newcommand{\ba}{\begin{array}} 
\newcommand{\ea}{\end{array}}

\newcommand{\lsim}{
\mathrel{\hbox{\rlap{\hbox{\lower4pt\hbox{$\sim$}}}\hbox{$<$}}}}
\newcommand{\gsim}{
\mathrel{\hbox{\rlap{\hbox{\lower4pt\hbox{$\sim$}}}\hbox{$>$}}}}

\newcommand{\newc}{\newcommand}

\newc{\beq}{\begin{equation}}
\newc{\eeq}{\end{equation}}
\newc{\beqn}{\begin{eqnarray}}
\newc{\eeqn}{\end{eqnarray}}
\newc{\bsym}{\boldsymbol}

\title{Discrete gauge symmetries and proton stability\\in the $\boldsymbol{U(1)'}$-extended MSSM}

\author{Hye-Sung Lee\\
Institute for Fundamental Theory, University of
Florida, Gainesville, FL 32611, USA\\
E-mail: \email{hlee@phys.ufl.edu}}
\author{Christoph Luhn\\
Institute for Fundamental Theory, University of
Florida, Gainesville, FL 32611, USA\\
E-mail: \email{luhn@phys.ufl.edu}}
\author{Konstantin T. Matchev\\
Institute for Fundamental Theory, University of
Florida, Gainesville, FL 32611, USA\\
E-mail: \email{matchev@phys.ufl.edu}}

\preprint{UFIFT-HEP-07-19}

\abstract{The Minimal Supersymmetric Standard Model (MSSM) with conserved
$R$-parity  suffers from several fine-tuning problems, e.g. the
$\mu$-problem and  the problem of proton decay through higher dimension
operators. Both of these problems can be avoided by replacing 
$R$-parity with a non-anomalous $U(1)'$ gauge symmetry 
which is broken at the TeV scale. The new gauge symmetry
does not necessarily forbid {\em all} renormalizable $R$-parity violating
interactions among the MSSM fields, and may allow for either lepton 
number or baryon number violation at the renormalizable level. 
However, the proton decay problem resurfaces with the introduction of
new TeV-scale exotic fields which are required for gauge anomaly cancellations.
In this paper we investigate the issue of proton stability in the presence of 
TeV-scale exotics. We show that there are large classes of models in which
TeV exotics do {\em not} destabilize the proton. We classify the viable models
according to the residual discrete symmetries after $U(1)'$ and electroweak
symmetry breaking. In some of our examples the residual $U(1)'$ discrete 
gauge symmetry within the MSSM sector alone ensures that the proton is 
absolutely stable, for {\em any} exotic representations. In other cases the
proton can be sufficiently long-lived, depending on the $U(1)'$ and hypercharge 
discrete charge assignments for the exotic fields. Our analysis outlines
a general scheme for ensuring proton stability in the presence of light 
exotics with baryon and lepton number violating interactions.}

\keywords{Discrete and Finite Symmetries, Beyond Standard Model, Supersymmetry Phenomenology}

\begin{document}

%%%%%%%%%%%%%%%%%%%%%%%%%%%%%%%%%%%%%%%%%%%%%%%%%%%%%%%%%%%
\section{Introduction}
%%%%%%%%%%%%%%%%%%%%%%%%%%%%%%%%%%%%%%%%%%%%%%%%%%%%%%%%%%%
In the Standard Model (SM)
lepton number ($\mathcal L$) and baryon number ($\mathcal B$) are conserved 
at the renormalizable level due to accidental global symmetries.
In the supersymmetric SM, with the addition of the superpartners, 
$\mathcal L$ and $\mathcal B$ are not conserved anymore.
Therefore, the supersymmetrization of the SM requires an accompanying 
symmetry or some mechanism for ensuring proton stability.
The Minimal version of the Supersymmetric Standard Model (MSSM)
with $R$-parity has been the most popular model of low-energy supersymmetry.
$R$-parity is a $\mathds Z_2$ symmetry, which has been the 
prevailing candidate for the companion symmetry of supersymmetry 
as it protects the proton from decaying through {\em renormalizable}
lepton number violating (LV) and baryon number violating (BV) terms.

However, $R$-parity alone does not completely cure 
the fine-tuning problems of the supersymmetric SM.
First, $R$-parity still allows the existence of dangerous higher dimension 
operators (e.g. $QQQL$ and $U^cU^cD^cE^c$ in the superpotential),
which violate both $\mathcal L$ and $\mathcal B$ and thus endanger
proton stability \cite{Weinberg:1981wj,Ellis:1982wr,Ellis:1983qm,Harnik:2004yp}\footnote{See, 
e.g.~Refs.~\cite{Kurosawa:2001iq,Sayre:2006en,Mohapatra:2007vd}, to see how 
the problem can be alleviated in grand unified theories.}.
This should be considered a serious flaw of the MSSM, given that $R$-parity 
was introduced to ensure proton stability in the first place.
In addition, $R$-parity does not address the $\mu$-problem \cite{Kim:1983dt} 
of the MSSM, whose solution may require some other mechanism.
These shortcomings of the MSSM motivate an alternative supersymmetrization
of the SM and/or an alternative companion symmetry, especially since
$R$-parity violation (RPV) leads to interesting phenomenology 
which is in agreement with all current experimental 
constraints \cite{Barbier:1998fe,Barbier:2004ez,Barger:1989rk,Allanach:2007vi,RPVcollider}.

The $U(1)'$-extended MSSM (UMSSM) \cite{UMSSM} is an extension of the 
MSSM with a new Abelian non-anomalous gauge symmetry $U(1)'$ 
at the TeV scale\footnote{Supersymmetric RPV models 
with an additional anomaly-free $U(1)$ gauge symmetry
have previously been considered in \cite{RPVU1prime}.  
For anomalous $U(1)$ approaches, see for example 
Ref.~\cite{Leontaris:1999wf} and references therein.}. 
In the UMSSM the $\mu$-problem is solved by replacing 
the original $\mu$ term ($H_2 H_1$) with an effective $\mu$ term ($S H_2 H_1$)
in the superpotential. Interestingly, 
it was recently found that the set of $U(1)'$ charge assignments 
which solve the $\mu$-problem, {\em automatically} forbid the 
coexistence of the renormalizable LV terms and BV terms, a phenomenon
which was dubbed {\em LV-BV separation} \cite{Lee:2007fw}.
Furthermore, the $U(1)'$ symmetry also guarantees the absence of dangerous
non-renormalizable proton decay operators constructed out of MSSM fields. 
Thus the $U(1)'$ symmetry ties up the explanation of the proton 
longevity to the solution to the $\mu$-problem and provides a solid 
theoretical framework for RPV phenomenology.  We therefore find
the $R$-parity violating UMSSM 
worth investigating as an alternative to the usual $R$-parity conserving MSSM. 

However, the new gauge symmetry usually requires some exotic fields at 
the $U(1)'$ breaking scale, in order to cancel the gauge 
anomalies \cite{CDM,Erler,AokiOshimo,Ma:2002tc}.
Such light exotics would reintroduce the proton stability problem, 
since the exotics themselves may have LV and/or BV interactions, and
may correspondingly mediate proton decay at unacceptable rates.
Therefore, the argument for proton stability in the UMSSM needs to
be extended to include the exotic representations. 
This discussion was postponed in Ref.~\cite{Lee:2007fw} and we shall complete it here.
We shall systematically study the proton decay problem in the UMSSM
by identifying the underlying discrete symmetries encoded in the 
set of phenomenologically viable $U(1)'$ charge assignments. 
We shall then use the $U(1)'$ discrete symmetry to argue that the 
proton is sufficiently stable even in the presence of light exotics.
For simplicity we shall mostly concentrate on $\mathds Z_3$ symmetries,
although we shall consider more general $\mathds Z_N$ examples as well.

Our UMSSM setup is reviewed in Section~\ref{framework}, and in
Section~\ref{dis} we identify the possible $U(1)'$ discrete 
gauge symmetries $\mathds Z_N$ among the MSSM fields only.
Of special interest to us will be the three $\mathds Z_3$ symmetries
denoted as $B_3$, $L_3$ and $M_3$ (see Section~\ref{dis} for their exact definition).
In the case of $B_3$, the $U(1)'$ discrete gauge symmetry among the MSSM fields 
is already sufficient to argue for the {\em absolute} stability of the proton
(see Section~\ref{absolute}). In case of $L_3$ and $M_3$,
the $U(1)'$ discrete symmetry
needs to be extended to include the exotics fields as well
(see Section~\ref{disexotics}) and the analysis becomes more involved.
Nevertheless, we still find various classes of models in which the proton
lifetime is sufficiently long. Our argument is based on the combination of the 
$U(1)'$ discrete gauge symmetry $\mathds Z_N$ {\em and} 
the hypercharge discrete gauge symmetry $\mathds Z_N^Y$
which is left over after electroweak symmetry breaking. In Section~\ref{good},
we identify all such ``good'' classes of models for the case of
$\mathds Z_3^Y \times \mathds Z_3$ discrete symmetries. The corresponding
results for the $\mathds Z_3^Y \times \mathds Z_6$-type extensions are 
presented in Appendix~\ref{app-z6}.
Sections~\ref{l3symmetric} and~\ref{z3yzn} provide explicit examples of
anomaly-free $U(1)'$ models. These serve the purpose of illustrating the
successive steps which are necessary to check that the proton is sufficiently
stable within a given model. Section~\ref{l3symmetric} showcases all
viable $L_3$ symmetric models with a $\mathds Z_3^Y \times \mathds Z_3$
extended discrete symmetry. The $L_3$ symmetric models of
Ref.~\cite{Lee:2007fw} with a $\mathds Z_3^Y \times \mathds Z_{N>3}$
extension are discussed in Section~\ref{z3yzn}, completing the proof of 
the claim made in Ref.~\cite{Lee:2007fw} that the proton is sufficiently stable
in these models. In addition, we also present some examples of
$M_3$ symmetric $\mathds Z_3^Y \times \mathds Z_{N>3}$-type charge assignments.
Section~\ref{conclusion} summarizes our results.

%%%%%%%%%%%%%%%%%%%%%%%%%%%%%%%%%%%%%%%%%%%%%%%%%%%%%%%%%%%
\section{The Framework of the UMSSM}
\label{framework}
%%%%%%%%%%%%%%%%%%%%%%%%%%%%%%%%%%%%%%%%%%%%%%%%%%%%%%%%%%%
In this section we briefly review the framework of the $U(1)'$-extended
Minimal Supersymmetric Standard Model. We closely follow the conventions 
in Ref.~\cite{Lee:2007fw}. In order to break the $U(1)'$ gauge
symmetry spontaneously, we need to introduce a Higgs singlet $S$ in addition to
the MSSM fields. $S$ is a singlet under the SM gauge group, but carries non-zero $U(1)'$ charge. 
To solve the $\mu$-problem, we require the $U(1)'$ charges to be such that 
the original $\mu$ term is forbidden but an effective $\mu$ term is 
dynamically generated after $S$ acquires a vacuum expectation value (vev) 
at the TeV scale. Accordingly, we require the superpotential term
\beq
W_{\mu} ~ = ~ h S H_2 H_1 \ .\label{eqmu}
\eeq
With regard to the Yukawa interactions, we assume 
\beq
W_{\mathrm{Yukawa}} ~=~ y^U_{jk} H_2 Q_j U^c_k +  y^D_{jk} H_1 Q_j D^c_k 
+ y^E_{jk} H_1 L_j E^c_k + y^N_{jk} \left(\frac{S}{M}\right)^a H_2 L_j N^c_k\ ,
\label{eqyuk}
\eeq
where we apply the standard notation for the MSSM superfields. The indices
$j,k$ label the three generations of quarks and leptons. 
Notice that we have also included 
Yukawa couplings for the right-handed neutrinos $N^c$. 
To explain the small neutrino mass in the case of pure Dirac 
neutrinos \cite{Langacker:1998ut}, we promote the Yukawa coefficient 
for the last term to have a suppression factor $\left(\frac{S}{M}\right)^a$ 
with a cutoff scale $M$ and a positive-definite integer $a$.\footnote{The 
exact numerical value of $M$ is not crucial for our discussion below. 
For example, $M$ could be taken as high as the Planck scale, or as low as 
$\mathcal O (10^{15}~ \text{GeV})$, as required for pure Dirac 
neutrino masses for $a=1$ and $y^N \sim \mathcal O (1)$.}

Assuming generation independence\footnote{Generation-dependent $U(1)'$ 
charges may explain the discrepancies in rare $B$ decays \cite{Langacker:2000ju,Bpuzzles}.}, 
eqs.~(\ref{eqmu}) and (\ref{eqyuk}) yield five constraints on nine 
$U(1)'$ charges. Denoting these charges by $z$, we have
\beqn
Y_S &:& z[S] + z[H_1] + z[H_2] = 0 \ ,\\
Y_U &:& z[H_2] + z[Q] + z[U^c] = 0 \ ,\\
Y_D &:& z[H_1] + z[Q] + z[D^c] = 0 \ ,\\
Y_E &:& z[H_1] + z[L] + z[E^c] = 0 \ ,\\
Y_N &:& z[H_2] + z[L] + z[N^c] + a z[S] = 0 \ .
\eeqn
With the above relations, the $[SU(3)_C]^2-U(1)'$ anomaly $A_{331'}$ cannot
vanish unless we introduce additional colored particles. A minimal\footnote{The only 
other alternative which would make $A_{331'}$ vanish, is an octet under $SU(3)_C$
\cite{CDM}. However, the choice of an octet is inconsistent with the remaining
anomaly conditions, unless one is willing to unnecessarily complicate the exotic 
particle spectrum.} and commonly considered extension 
of the particle spectrum is to assume three generations \cite{Lee:2007fw} 
of exotic quarks $K_i$, which are triplets under $SU(3)_C$ and singlets 
under $SU(2)_L$, as well as their right-handed counterpartners $K^c_i$. 
These exotic fields acquire their masses at the scale of 
$U(1)'$ breaking from the superpotential terms 
\beq
W_{\mathrm{exotic}} ~=~ h_{ij} S K_i K^c_j \ ,
\label{exmass}
\eeq
which we assume to have non-vanishing diagonal couplings. Then, the $U(1)_Y \times
U(1)'$ charges of the $K^c_i$ are uniquely fixed by those of the $K_i$
\beq
y[K^c_i] ~=~ - y[K_i]\ , \qquad z[K^c_i] ~=~ - z[K_i] - z[S] \ .
\eeq
While canceling $A_{331'}$, the exotic quarks introduce six new parameters. 
However, they do not affect the $[SU(2)_L]^2-U(1)'$ anomaly $A_{221'}$
\beq
A_{221'} ~:~ 3 (3z[Q] + z[L]) + N_H (z[H_1] + z[H_2]) = 0\ .
\label{A221}
\eeq
Eq.~(\ref{A221}) yields another constraint on the $U(1)'$
charges. Setting aside the exotic quarks for the moment, 
we can express the $U(1)'$ charges of the remaining fields
in terms of three free parameters $\alpha$, $\beta$ and $\gamma$:
\beq
\left(\begin{array}{l} z[Q] \\ z[U^c] \\ z[D^c] \\ z[L] \\ z[N^c] \\ z[E^c] 
                \\ z[H_1] \\ z[H_2] \\ z[S] \end{array}\right)
~=~ \alpha 
\left(\begin{array}{r} 1\\-4\\2\\-3\\0\\6\\-3\\3\\0   \end{array}\right)
~+~ \beta
\left(\begin{array}{r} 1\\-1\\-1\\-3\\3\\3\\0\\0\\0   \end{array}\right)
~+~ \gamma 
\left(\begin{array}{r} N_H\\9-N_H\\-N_H\\0\\9(1-a)\\0\\0\\-9\\9 \end{array}\right)  \ .\label{master}
\eeq
The first vector is proportional to the SM hypercharge, and the second 
corresponds to ${\mathcal B} - {\mathcal L}$. One can therefore
use the freedom contained in the second and third vector to allow 
or forbid SM-invariant operators in the superpotential and the 
K\"ahler potential.

Since we do not a priori assume $R$-parity, the usual $R$-parity violating 
terms are allowed in general:
\beqn
W_{\mathrm{LV}} &=& \hat \mu_i L_iH_2  
~+~ \hat \lambda_{ijk} L_iL_jE^c_k
~+~ \hat \lambda'_{ijk} L_iQ_jD^c_k \ ,  \label{WLVfirst}\\
W_{\mathrm{BV}} &=& \hat\lambda''_{ijk} U^c_iD^c_jD^c_k \ .\label{WBVfirst}
\eeqn
Motivated by the relatively tight experimental constraints 
on the individual RPV couplings $\hat\mu$, $\hat\lambda$, $\hat\lambda'$ 
and $\hat\lambda''$ \cite{Barbier:2004ez}, 
we shall exploit the possibility that they may originate
from higher dimension operators, and their values are suppressed by
factors of $\left(\frac{\phi}{M}\right)^\ell$ where $\phi$ is a 
SM singlet combination of fields which acquires a vev, in our case $S$ or 
$H_2 H_1$. In general, we then have
\beq
\hat\lambda = \lambda \left( \frac{\left<S\right>}{M} \right)^A 
\left( \frac{\left<H_2\right> \left<H_1\right>}{M^2} \right)^B \ ,
\label{ABsup}
\eeq
(with positive-definite $A$ and $B$) and similarly for $\hat\mu$,
$\hat\lambda'$ and $\hat\lambda''$.
For the sake of simplicity, in what follows we shall typically assume that $B=0$,  
so that all suppression factors are of the type $\left(\frac{\left<S\right>}{M}\right)^A$.
This assumption is not crucial to our discussion, and only in
Sections~\ref{l3symmetric} and \ref{z3yzn} we shall revisit this issue, 
allowing for $\left(\frac{\left<H_2\right>\left<H_1\right>}{M^2}\right)^B$
type of suppression as well.

With those assumptions, the corresponding RPV superpotentials
(\ref{WLVfirst}) and (\ref{WBVfirst}) become
\beqn
W_{\mathrm{LV}} &=& h'_i \left(\frac{S}{M} \right)^n SL_iH_2  
~+~ \lambda_{ijk} \left(\frac{S}{M} \right)^{n} L_iL_jE^c_k
~+~ \lambda'_{ijk} \left(\frac{S}{M} \right)^{n} L_iQ_jD^c_k \ , \label{eq:WLV} \\
W_{\mathrm{BV}} &=& \lambda''_{ijk} \left(\frac{S}{M}\right)^m U^c_iD^c_jD^c_k \ .
\label{eq:WBV}
\eeqn
Notice that all three LV terms in (\ref{eq:WLV}) have the same power of $1/M$ 
suppression, for which from now on we shall use $n$, while for the
corresponding suppression in (\ref{eq:WBV}) we shall use $m$.
The integers $n$ and $m$ should be considered among the input parameters 
of our UMSSM models.

Following Ref.~\cite{Lee:2007fw}, we can easily demonstrate the
LV-BV separation by taking the linear combination 
$6Y_D + 3Y_U - 3Y_E + (N_H-3) Y_S - A_{221'}$, resulting in
\beq
3z[S^m U^c D^c D^c] - 3z[S^n L L E^c] + (N_H - 3(1+m-n))z[S] = 0 \ . \label{eq:LVBVseparation}
\eeq
We see that the LV-BV separation observed in Ref.~\cite{Lee:2007fw}
can now be generalized for any values of $n$ and $m$. 
As long as the condition $N_H \ne 3 \cdot \mathbb Z$ 
is kept, the third term in eq.~(\ref{eq:LVBVseparation}) does not vanish
and must be canceled by one (or the combination) of the first two terms.
Notice that the first (second) term in eq.~(\ref{eq:LVBVseparation})
is nothing but the $U(1)'$ charge of the BV (LV) operator(s) in
eq.~(\ref{eq:WBV}) (eq.~(\ref{eq:WLV})). Therefore, the nonvanishing 
of the first or second term in eq.~(\ref{eq:LVBVseparation}) implies that
the corresponding renormalizable RPV couplings (BV or LV) are absent
from the superpotential, as they are forbidden by the $U(1)'$ gauge symmetry. 
The fact that certain terms are forbidden even at the non-renormalizable level 
(i.e. with arbitrary suppression factors $(S/M)^\ell$) suggests 
a certain symmetry. In Section \ref{dis} we shall investigate 
the nature of the symmetry which is implied by the phenomenon of 
LV-BV separation. In what follows we shall restrict ourselves to 
the simplest and most natural case exhibiting LV-BV separation, 
namely $N_H = 1$.

The additional requirement of having either the LV terms 
$LH_2$, $LLE^c$, $LQD^c$ of eq.~(\ref{eq:WLV}) or the BV terms 
$U^cD^cD^c$ of eq.~(\ref{eq:WBV}) at the effective level reduces 
the general solution eq.~(\ref{master}) to a two-parameter solution.
\begin{itemize}
\item For the LV case, one must demand that
\beq
z[H_1]~=~z[L]+nz[S]\ ,\label{lvcond}
\eeq
from eq.~(\ref{eq:WLV}).
Eq.~(\ref{lvcond}) relates the parameters $\beta$ and $\gamma$ in eq.~(\ref{master}) 
by the condition $\beta = 3 n \gamma$. The $U(1)'$ charges
in the LV case can then be written as
\beq
\left(\begin{array}{l} z[Q] \\ z[U^c] \\ z[D^c] \\ z[L] \\ z[N^c] \\ z[E^c] 
                \\ z[H_1] \\ z[H_2] \\ z[S] \end{array}\right)
~=~ \left( \alpha+(3n+N_H)\gamma \right) 
\left(\begin{array}{r} 1\\-4\\2\\-3\\0\\6\\-3\\3\\0   \end{array}\right)
~+~ 3 \gamma
\left(\begin{array}{r}
    0\\3(1+n)+N_H\\-3n-N_H\\N_H\\3(1-a+n)\\-3n-2N_H\\3n+N_H\\-3(1+n)-N_H\\3
  \end{array}\right) \ .\label{LVmaster}
\eeq
\item In the BV case, we must require
\beq
z[H_1]~=~z[L]+\left(1+m-\frac{N_H}{3}\right)z[S]\ ,\label{bvcond}
\eeq
from eq.~(\ref{eq:WBV}).
$\beta$ and $\gamma$ are now related by $\beta = (3+3m-N_H) \gamma$, and we obtain
\beq
\left(\begin{array}{l} z[Q] \\ z[U^c] \\ z[D^c] \\ z[L] \\ z[N^c] \\ z[E^c] 
                \\ z[H_1] \\ z[H_2] \\ z[S] \end{array}\right)
~=~ \left( \alpha+3(1+m)\gamma \right)
\left(\begin{array}{r} 1\\-4\\2\\-3\\0\\6\\-3\\3\\0   \end{array}\right)
~+~ 3 \gamma
\left(\begin{array}{r}
    0\\3(2+m)\\-3(1+m)\\N_H\\3(2-a+m)-N_H\\-3(1+m)-N_H\\3(1+m)\\-3(2+m)\\3
  \end{array}\right) \ .\label{BVmaster}
\eeq
\end{itemize}

%%%%%%%%%%%%%%%%%%%%%%%%%%%%%%%%%%%%%%%%%%%%%%%%%%%%%%%%%%%
\section{Discrete Symmetries Without Exotics}
\label{dis}
%%%%%%%%%%%%%%%%%%%%%%%%%%%%%%%%%%%%%%%%%%%%%%%%%%%%%%%%%%%
Within the framework of the UMSSM, there are the usual MSSM particles, plus the
right-handed neutrinos~$N^c_i$, the Higgs singlet $S$ and
the exotic quarks~$K_i$, $i=1,2,3$. First, we want to discuss the occurrence of discrete
symmetries at the {\it effective level}, where the $K_i$
are integrated out. In that case, a general superpotential or K\"ahler
potential operator has $n_Q$ quark doublets, $n_{U^c}$ up-type antiquark
singlets, etc. If all $n_{...}$ are positive, the corresponding term 
appears in the superpotential. Negative $n_{...}$ are used for
conjugate superfields, thus an operator with some $n_{...}$ being positive 
and others negative, can only appear in the K\"ahler potential. 
SM gauge invariance requires a certain relation among the $n_{...}$, see e.g.
Refs.~\cite{Dreiner:2003yr,Dreiner:2006xw}. The total $U(1)'$ charge of such a
generic operator without exotic quarks is given as
\beqn
z[\mathrm{op.}]  
&=&  n_Q z[Q] + n_{U^c} z[U^c] + n_{D^c} z[D^c] + n_L z[L] + n_{N^c} z[N^c]
+ n_{E^c} z[E^c] \nonumber \\
&&  + n_{H_1} z[H_1] + n_{H_2} z[H_2] + n_S z[S] \ . \label{totalz}
\eeqn
The transition from the continuous $U(1)'$ to its discrete subgroup $\mathds
Z_N$ is made by choosing a normalization in which the $U(1)'$ charges are all
integers, with 
\beq
|z[S]| \equiv N \ , 
\eeq
and then defining the discrete charge $q[F] \in [0,N-1]$ of a field $F$ by the
mod~$N$ part of the corresponding $U(1)'$ charge
\beq
z[F] = q[F] +  N\cdot \mathbb Z \ .\label{defq}
\eeq
Plugging this into eq.~(\ref{totalz}), we can define the discrete charge of
any superpotential or K\"ahler potential operator by
\beqn
q[\mathrm{op.}]&\equiv&  n_Q q[Q] + n_{U^c} q[U^c] +  n_{D^c} q[D^c]  +  n_L
q[L]  + n_{N^c} q[N^c] + n_{E^c} q[E^c] \nonumber \\
&&  + n_{H_1} q[H_1] +   n_{H_2} q[H_2] \ .
\eeqn
The $\mathds Z_N$ symmetry forbids any operator whose discrete charge
$q[\mathrm{op.}]$ is not a multiple of~$N$. If it is a multiple of~$N$, the
operator might or might not exist, depending on the actual $U(1)'$
charges. At this stage, we only want to argue from the discrete symmetry
viewpoint. 

The discrete anomaly-free $\mathds Z_N$ symmetries of the MSSM fields with and without right-handed
neutrinos $N^c$ are known \cite{Ibanez:1991hv,Ibanez:1991pr,Dreiner:2005rd,Luhn:2007gq}.
For $N=3$, for example, all possibilities are shown in Table~\ref{table1}.
$B_3$ is the well-known baryon triality, and $L_3$ is correspondingly the lepton triality.
$R_3$ is a symmetry related to the right-handed isospin. Finally,
we have defined matter triality $M_3 \equiv R^{}_3L^2_3$.
We point out that an overall sign flip of the discrete charges 
leads to exactly the same discrete symmetries. Such a sign
change amounts to the exchange $1 \leftrightarrow 2$ in Table~\ref{table1}.

\TABLE[t]{
\begin{tabular}{|c||c|c|c|c|c|c|c|c|} \hline
$\phantom{\Big|}$ & $q[Q]$ & $q[U^c]$ & $q[D^c]$ & $q[L]$ & $q[N^c]$ & $q[E^c]$ & $q[H_1]$ & $q[H_2]$ \\ \hline\hline
$B_3$ & $0$ & $2$ & $1$ & $2$ & $0$ & $2$ & $2$ & $\phantom{\Big|}1$ \\ \hline 
$L_3$ & $0$ & $0$ & $0$ & $2$ & $1$ & $1$ & $0$ & $\phantom{\Big|}0$ \\ \hline 
$M_3$ & $0$ & $2$ & $1$ & $1$ & $1$ & $0$ & $2$ & $\phantom{\Big|}1$ \\ \hline 
$R_3$ & $0$ & $2$ & $1$ & $0$ & $2$ & $1$ & $2$ & $\phantom{\Big|}1$ \\ \hline 
\end{tabular}\caption{\label{table1} The discrete anomaly-free $\mathds Z_3$
  symmetries defined on the MSSM+$N^c$ sector.}
}

We now extract the discrete symmetries which are encoded in the
$U(1)'$ charges of eqs.~(\ref{LVmaster}) and (\ref{BVmaster}), i.e. the
discrete symmetries for the LV case and the BV case. First, note that the hypercharge
vector is irrelevant for the discussion of the remnant discrete symmetry of
the $U(1)'$ since the discrete symmetries are equivalent up to a 
shift by the hypercharges (normalized to integers).
So it suffices to focus on the second vector (proportional to $3 \gamma$) of
eqs.~(\ref{LVmaster}) and (\ref{BVmaster}). In both cases, 
the minimum $N$ is $N=z[S]=3$ with $3 \gamma = 1$.
The discrete symmetries should be $\mathds Z_3$ for the LV and the BV cases.
More explicitly, we find
\beq
\left(\begin{array}{l} q[Q] \\ q[U^c] \\ q[D^c] \\ q[L] \\ q[N^c] \\ q[E^c] 
                \\ q[H_1] \\ q[H_2]  \end{array}\right)_{\mathrm{LV}}
=~~
\left(\begin{array}{r}
    0\\N_H\\2N_H\\N_H\\0\\N_H\\N_H\\2N_H
  \end{array}\right) \ ,
\qquad\qquad
\left(\begin{array}{l} q[Q] \\ q[U^c] \\ q[D^c] \\ q[L] \\ q[N^c] \\ q[E^c] 
                \\ q[H_1] \\ q[H_2]  \end{array}\right)_{\mathrm{BV}}
=~~
\left(\begin{array}{r}
    0\\0\\0\\N_H\\2N_H\\2N_H\\0\\0
  \end{array}\right) \ ,
\eeq
where we have used that $\mp N_H= \pm 2 N_H~\mbox{mod}~3$.
Note that these results do not depend on the specific values for 
$n$ or $m$. Comparing with
Table~\ref{table1} shows that, for $N_H \neq 0~\mbox{mod}~3$, the LV case
yields baryon triality $B_3$ whereas the BV case gives rise to lepton
triality $L_3$. Therefore, the origin of the LV-BV separation
exhibited by eq.~(\ref{eq:LVBVseparation}) can now be traced 
back to the existence of the discrete symmetries $L_3$ and $B_3$. 

So far, we have investigated the discrete symmetries encoded in the UMSSM 
which solves the $\mu$-problem. We identified $B_3$
in the LV case (where the $LH_2$, $LLE^c$ and 
$LQD^c$ terms are effectively present), 
and $L_3$ in the BV case (where the $U^cD^cD^c$ term 
effectively appears). Of course, 
one does not have to require any of these lepton or baryon number
violating interactions. Then one can end up with other discrete symmetries as
well. To see this, we rewrite eq.~(\ref{master}) by constructing a new basis 
of the three-parameter solution in which the first component of both the
second and the third vector is zero. 
With $\alpha' = \alpha + \beta + N_H \gamma$, $\beta' = 3 \beta$, $\gamma' = 3 \gamma$, we get
\beq
\left(\begin{array}{l} z[Q] \\ z[U^c] \\ z[D^c] \\ z[L] \\ z[N^c] \\ z[E^c] 
                \\ z[H_1] \\ z[H_2] \\ z[S] \end{array}\right)
~=~ \alpha' 
\left(\begin{array}{r} 1\\-4\\2\\-3\\0\\6\\-3\\3\\0   \end{array}\right)
~+~ \beta'
\left(\begin{array}{r} 0\\1\\-1\\0\\1\\-1\\1\\-1\\0   \end{array}\right)
~+~ \gamma' 
\left(\begin{array}{r} 0\\3+N_H\\-N_H\\N_H\\3(1-a)\\-2N_H\\N_H\\-3-N_H\\3 \end{array}\right)  \ .\label{master'}
\eeq
Assuming $N_H=1$, we obtain matter triality $M_3$ as the remnant discrete
symmetry in models where $\beta'=1\,\mbox{mod}\,3$ and $\gamma'=1$ 
(or $\beta'=-1\,\mbox{mod}\,3$ and $\gamma'=-1$). 
It is worth pointing out that the charge assignment of $R_3$ has $q[L]=0$,
which, due to our previous assumption $N_H \neq 0~\mathrm{mod}~3$, 
can only be obtained for $\gamma'=0$. This, however, is inconsistent,
as the Higgs singlet $S$ would then be neutral under $U(1)'$. 
For $N_H=1$ there are thus only three $\mathds Z_3$ symmetries 
which can be generated from eq.~(\ref{master'}):
\beq
\begin{array}{rll}
B_3:~~~~ & \beta'= 0~\mbox{mod}~3 \ ,~~~ & \gamma'= 1 \ , \\ %\\[1mm]
M_3:~~~~ & \beta'= 1~\mbox{mod}~3 \ ,~~~ & \gamma'= 1 \ , \\ %\\[1mm]
L_3:~~~~ & \beta'= 2~\mbox{mod}~3 \ ,~~~ & \gamma'= 1 \ . \\ %\\[1mm]
\end{array}
\label{eq:Z3symmetries}
\eeq
With $\gamma' = -1$, the above $\beta'$ would also have to flip sign in order to yield 
the same discrete symmetries. Choosing $\gamma' \neq \pm 1$ 
generically leads to $\mathds Z_N$ symmetries with higher $N$.
However, notice that the LV case (i.e. with $LLE^c$ etc.) 
and the BV case ($U^cD^cD^c$) requirement leads only to 
$B_3$ and $L_3$ or their simple scaling (such as $B_6^2$, $L_6^2$), 
respectively. 

The discrete $\mathds Z_N$ symmetries encoded in the $U(1)'$ charges provide
a powerful tool to see which $\mathcal L$ and/or $\mathcal B$ violating operators
could in principle arise in the theory. All such operators up to dimension five 
are summarized in Table~\ref{table2} for the three possible $\mathds Z_3$ 
symmetries in eq.~(\ref{eq:Z3symmetries}).
\TABLE[t]{
\begin{tabular}{|c||c|c|} \hline
&\begin{tabular}{c} operators with \\ $\mathcal B$ violation \end{tabular}   
&\begin{tabular}{c} operators with \\ $\mathcal L$ violation \end{tabular}
\\ \hline\hline
$B_3$ & none & $\begin{array}{c}  
N^c; \\
LH_2,~  N^cN^c,~ N^cS; \\
LQD^c,~ LLE,~ SLH_2,~ N^cH_1H_2,~ LH_1^\dagger,\\
N^cN^cN^c,~  N^cN^cS,~  N^cSS,~  N^cS^\dagger;~ \\ 
\mbox{many~dimension~five~terms}
\end{array}$ \\ \hline
$L_3$ & 
$\begin{array}{c} 
U^cD^cD^c; \\
QQQH_1,~ SU^cD^cD^c,~ QQ{D^c}^\dagger
\end{array}$ &
$\begin{array}{c} 
N^cN^cN^c; \\
SN^cN^cN^c
\end{array}$ \\ \hline
$M_3$ & none & 
$\begin{array}{c} 
N^cN^cN^c; \\
SN^cN^cN^c
\end{array}$ \\ \hline
\end{tabular}\caption{\label{table2}$\mathcal B$ and/or $\mathcal L$ violating
operators up to dimension five which conserve a $\mathds Z_3$ symmetry and
comprise only MSSM particles, right-handed neutrinos $N^c$ and Higgs singlets
$S$.}}
As Table \ref{table2} demonstrates, $B_3$ allows a number of LV terms 
but does not allow the BV terms of eq.~(\ref{eq:WBV}), in accord with LV-BV 
separation. Similarly, $L_3$ allows BV terms but does not allow 
the LV terms of eq.~(\ref{eq:WLV}).
Finally, $M_3$ forbids both the LV terms of eq.~(\ref{eq:WLV}) and 
the BV terms of eq.~(\ref{eq:WBV}).

Since proton decay requires both baryon number violation as well as lepton
number violation, the absence of either one of them is sufficient to stabilize the
proton. Moreover, any $\mathcal B$ and/or $\mathcal L$ violating interaction which
is suppressed by two powers of a high cutoff scale $M$ does not
endanger the proton.  That is why Table~\ref{table2} lists only
superpotential and K\"ahler potential operators up to dimension five.
$B_3$ and $M_3$ conserve $\mathcal B$ up to this level, while $L_3$ and $M_3$
have $\mathcal L$ violation only through the two operators $N^cN^cN^c$ and 
$SN^cN^cN^c$, which could be forbidden by a judicious choice of
the $U(1)'$ charges $z[N^c]$ and $z[S]$. Table~\ref{table2} 
reveals that in models exhibiting a $B_3$ or $M_3$ discrete symmetry,
the proton cannot be destabilized by {\em any} diagram involving
MSSM fields, $N^c$ and $S$. In models with an $L_3$ symmetry, 
it is simply sufficient that one forbids the $N^cN^cN^c$ and 
$SN^cN^cN^c$ superpotential terms, and the proton is safe from such diagrams 
as well.

While these statements are true to all orders in perturbation theory, 
they are of limited use due to an important caveat which we must take into account.
So far in this section we have ignored the effect of the exotics.
Once we take them into account, the scale which suppresses 
the non-renormalizable operators in Table~\ref{table2} 
is by far not guaranteed to be the high scale $M$.
In fact, the exotics are relatively light, near the TeV scale,
since they get their masses from the $U(1)'$ breaking scale (see eq.~(\ref{exmass})).
Thus, depending on their couplings to the MSSM+$N^c$ sector, 
the exotics could in principle destabilize the proton.

Before delving into the details of how to take account of the exotic quarks,
we shall take a detour to explain why certain $\mathds Z_N$ symmetries such as 
baryon triality $B_3$ or proton hexality $P_6$~\cite{Dreiner:2005rd} 
are actually sufficient to {\em completely} stabilize the
proton, independent of the presence of any light exotics or their properties.

%%%%%%%%%%%%%%%%%%%%%%%%%%%%%%%%%%%%%%%%%%%%%%%%%%%%%%%%%%%
\section{The Absolute Stability of the Proton}
\label{absolute}
%%%%%%%%%%%%%%%%%%%%%%%%%%%%%%%%%%%%%%%%%%%%%%%%%%%%%%%%%%%
We have just emphasized that proton decay requires violation of both $\mathcal B$ 
and $\mathcal L$. However, not any type of
$\mathcal B$ violation can lead to a decaying proton. In order to understand
this, first observe that the exotic quarks are heavier than the proton
(otherwise they would have been produced and seen at colliders). Hence, for
kinematic reasons, the proton cannot decay to exotic quarks in the final
state, and we can exclude the exotics $K_i$ from the discussion in this section.
Since the proton is the lightest particle with non-zero baryon number 
($\mathcal B=1$), the final state particles must be non-baryonic. Therefore, 
{\it proton decay requires an interaction which has $|\Delta \mathcal B| = 1$.} 
A theory  where $\mathcal B$ is only violated by a certain number of units
$\eta>1$ (so that {\em any} BV operator in the theory has $|\Delta \mathcal B|=0$ mod $\eta$),
automatically has an absolutely stable proton. This was observed in 
Ref.~\cite{Castano:1994ec} for the specific case of baryon triality $B_3$.

Here, we first review the argument for $B_3$, with the sign
convention adopted in Table~\ref{table1}. Since it is a $\mathds Z_3$
symmetry, the two Higgs doublets have opposite discrete charges
$-q[H_1]=q[H_2]=1~\mbox{mod}~3$. It is possible to redefine the $q[F]$ by
adding a certain amount (take e.g. $\alpha' =-\frac{1}{3}$) of the hypercharge
vector in eq.~(\ref{master'}), so that $q[H_1]=q[H_2]=0~\mbox{mod}~3$. 
As discrete charges should always be integer, it is necessary to rescale the
resulting vector by multiplication of 3. 
We obtain the $\mathds Z_9$ charge assignment, which is exactly $-3 \mathcal B$, 
as shown in the first row of Table~\ref{abs-table}.
Then the total discrete charge of any operator is
\beq
q[\mathrm{op.}] = -(n_Q - n_{U^c} - n_{D^c}) = 0 ~\mbox{mod}~9 \ , \label{word}
\eeq
while for proton decay we need $|\Delta \mathcal B| = 1$, i.e. 
\beq
n_Q-n_{U^c}-n_{D^c} = \pm 3 \ , \label{byone}
\eeq
which is incompatible with eq.~(\ref{word}).
Thus, the proton is absolutely stable if the discrete symmetry is $B_3$.

\TABLE[t]{
$
\begin{array}{|c||c|c|c|c|c|c|c|c|c|}\hline
\multicolumn{1}{|c||}{\phantom{\Big|}\mbox{discrete symmetry}\phantom{\Big|}} &
q[Q] & q[U^c] & q[D^c] &q[L] &q[N^c] &q[E^c] &q[H_1] &q[H_2] \\ \hline\hline 
B_3 \rightarrow \mathds Z_9 & -1 & 1 & 1 & 0&0&0&0&\phantom{\Big|}0 
\\ \hline
\mathds Z_T                    & t_Q&t_{U^c}&t_{D^c}&t_L&t_{N^c}&t_{E^c}&t_{H_1}&t_{H_2} \\ \hline
\mathds Z_9 \times \mathds Z_T & -T+9t_Q&T+9t_{U^c}&T+9t_{D^c}&9t_L&9t_{N^c}&9t_{E^c}&9t_{H_1}&9t_{H_2} \\ \hline
%P_6 & \rightarrow ~~ \mathds Z_{18} & -1 & 1 & 1 & 9&9&9&0&\phantom{\Big|}0 
%\\ \hline
%R^{}_{12}L^4_{12} & \rightarrow ~~ \mathds Z_{36} & -1 & 1 & 1 & -9~\,\;
%&9&9&0&\phantom{\Big|}0  \\ \hline
%R^{5}_{12}L^8_{12} & \rightarrow ~~ \mathds Z_{36} & -5 & 5 & 5 & -9~\,\;
%&9&9&0&\phantom{\Big|}0  \\ \hline
\end{array}
$
\caption{\label{abs-table}The discrete charges of the $B_3$ equivalent
  $\mathds Z_9$ and $\mathds Z_9 \times \mathds Z_T$.} }
This argument can be applied to more general cases.
If our $U(1)'$ has a discrete symmetry of $B_3 \times \mathds Z_T$ 
(with $T$ coprime to $3$), then the proton is also absolutely 
stable\footnote{The direct product of two cyclic groups 
($\mathds Z_A \times \mathds Z_B$) is a cyclic group $\mathds Z_{AB}$ 
if $A$ and $B$ have greatest common divisor $1$ (i.e. if they are coprime).}.
The resultant discrete charge of $\mathds Z_{9T}$ is given 
by $T q_{Z_9} + 9 q_{Z_T}$ as Table \ref{abs-table} shows.

There are examples of this kind in Ref.~\cite{Luhn:2007gq} where all 
anomaly-free $\mathds Z_{N\leq 14}$ symmetries were identified\footnote{See 
Ref.~\cite{Ibanez:1991pr} for the definition of the generators $R_N$ and
$L_N$ for arbitrary values of $N$.}: $R^{}_6L^4_6$ (proton hexality or 
$P_6$ \cite{Dreiner:2005rd}),  $R^{}_{12}L^4_{12}$ and $R^{5}_{12}L^8_{12}$. 
These three symmetries are isomorphic to the direct product of $B_3$ 
with some other $\mathds Z_T$ symmetries and provide absolute proton stability
\beqn
R^{}_6L^4_6        &\cong& B_3 \times R_2 \ , \\
R^{}_{12}L^4_{12}  &\cong& B_3 \times R_4^3 \ , \\
R^{5}_{12}L^8_{12} &\cong& B_3 \times R_4 \ .
\eeqn

%%%%%%%%%%%%%%%%%%%%%%%%%%%%%%%%%%%%%%%%%%%%%%%%%%%%%%%%%%%
\section{Discrete Symmetries with Exotic Quarks}
\label{disexotics}
%%%%%%%%%%%%%%%%%%%%%%%%%%%%%%%%%%%%%%%%%%%%%%%%%%%%%%%%%%%
In Table~\ref{table2}, we have listed only $\mathcal B$ and/or $\mathcal L$
violating operators up to dimension five. This is sufficient to argue for a
stable proton only under the assumption that the non-renormalizable
interactions between the particles are generated by high scale physics.  
However, since the exotic quarks were integrated out, the mass suppression of
the non-renormalizable terms could in principle be of $\mathcal
O(\mbox{TeV})$, the scale where the $U(1)'$ breaks down and gives mass to the
$K_i$. In such a case, one should also consider $\mathcal B$ and/or $\mathcal L$
violating operators with dimensionality higher than five. 

Alternatively, one can try to extend the $\mathds Z_N$ symmetries to
{\it explicitly include the exotic particles}. Under the
assumption that any  additional new physics (other than 
$U(1)'$) occurs only at the high scale, the non-renormalizable 
interactions involving the MSSM particles, the right-handed neutrinos, 
the Higgs singlet as well as the exotic quarks are all suppressed by powers of
$M$. Thus the advantage of this approach is that we still 
only need to consider a finite number of operators (up to dimension 5),
albeit over an extended set of fields.
Similarly to the case without exotic quarks, the obtained
{\em discrete} symmetries can be studied with respect to the allowed 
$\mathcal B$ and/or $\mathcal L$ violating operators, where
by definition the exotic quarks and antiquarks do not carry baryon 
or lepton number.

We now have to assign discrete charges to the three exotic quarks~$K_i$. 
The corresponding antiquarks $K^c_i$ automatically have the opposite 
discrete charges. Assuming that their original $U(1)'$ charges are integers, 
we have $27$ different $\mathds Z_3$
charge assignments for the exotics $K_i$. However, as their generations 
have not been defined yet, we need to consider only 10 of these 27
possibilities, see Table~\ref{table3}. 
\TABLE[t]{$~~~~~~~~~~~~
\begin{array}{|c||c|c|c|c|c|c|c|c|c|c|}\hline
\phantom{\Big|}q[K_1]\phantom{\Big|} &0~&~1~&~2~&~0~&~0~&~0~&~1~&~0~&~1~&~0~ 
 \\\hline
\phantom{\Big|}q[K_2]\phantom{\Big|} &~0~&~1~&~2~&~0~&~0~&~1~&~1~&~2~&~2~&~1~ 
 \\\hline
\phantom{\Big|}q[K_3]\phantom{\Big|} &~0~&~1~&~2~&~1~&~2~&~1~&~2~&~2~&~2~&~2~
 \\\hline
\end{array}
~~~~~~~~~~~~$\caption{\label{table3}10 of 27 possible $\mathds Z_3$ charge
assignments for the exotic quarks are sufficient since the labeling of the
generation for the exotics has not been determined yet.}}
When determining the allowed $\mathcal B$ and/or $\mathcal L$ violating
superpotential and K\"ahler potential operators in Section~\ref{dis}, we
have required invariance under the $\mathds Z_N$ subgroup of $U(1)'$, but also
- tentatively - SM gauge invariance. We do not know the hypercharges of the
$K_i$, and in principle there could be infinitely many hypercharge
assignments. In order to systematize this issue, we work in a normalization in
which $y[Q]=1$ and assume that the hypercharges of the exotic quarks are
integers. Now, we can  require invariance under any $\mathds Z^Y_N$ subgroup
of $U(1)_Y$, with the discrete hypercharges $q^Y$ being defined by the relation
\beq
y[F]=q^Y[F]+N\cdot \mathbb Z \ .
\eeq 
For simplicity, we choose\footnote{After
electroweak symmetry breaking (EWSB), the choice $N=3$ coincides with the
{\it remnant} discrete symmetry of $U(1)_Y$, as opposed to the
$\mathds Z^Y_N$ symmetries with arbitrary $N$ which exist before EWSB.}
$N=3$, so that we end up with $27$ discrete hypercharge assignments for the
exotic quarks. 

Starting from an $SU(3)_C \times SU(2)_L \times U(1)_Y \times U(1)'$ gauge
theory, we study only its subgroup $SU(3)_C \times SU(2)_L \times \mathds
Z^Y_3 \times \mathds Z_3$. The discrete charges of the MSSM particles are
uniquely defined once we have picked our preferred $\mathds Z_3$ symmetry
among the MSSM fields. Due to the different $U(1)_Y \times U(1)'$ charge
assignments for the exotic quarks, we have $10\times 27=270$ cases to study.
Each scenario is defined by the discrete hypercharges $q^Y[K_i]$ and the
discrete $U(1)'$ charges $q[K_i]$. We determine all $SU(3)_C \times SU(2)_L
\times \mathds Z^Y_3 \times \mathds Z_3$ invariant operators up to dimension
five and check whether there exists a conserved quantity~$\mathcal Q$ of the
form 
\beq
\mathcal Q ~ \equiv ~ b \, \mathcal B + \ell \, \mathcal L + k_1 \,
\mathcal K_1+ k_2 \, \mathcal K_2+ k_3 \, \mathcal K_3 \ ,\label{global}
\eeq
with $b,\ell,k_i$ being integers and $\mathcal K_i$ denoting (individual) exotic quark number,
respectively.  We stress that $\mathcal Q$ is conserved only among the
operators up to dimension five. However, since any operator that is
suppressed by two powers of $M$ is not dangerous for proton
decay, we loosely speak of ``$\mathcal Q$ conservation'' in the following,
mindful of its approximate meaning.   

Let us illustrate the implications of such a quantity $\mathcal Q$ with two 
examples. First, assume that the set of allowed operators up to dimension five
has $\mathcal Q_1 = \mathcal B$ (e.g. as model (v) of Table \ref{table7}).
Then, among these
operators, baryon number is conserved and the proton is sufficiently
stabilized. Next, we take $\mathcal Q_2 = 3\mathcal L-\mathcal K_1+2\mathcal
K_2+2\mathcal K_3$ (e.g. as model (i) of Table \ref{table7}).
In this case, both baryon and lepton number are
violated. Concerning proton decay, however, the exotic quarks $K_i$ are
heavier than the proton and therefore cannot be present among the final state
particles. So any diagram that potentially mediates proton decay necessarily
has $\mathcal K_i = 0$. Due to the conservation of~$\mathcal Q_2$, lepton
number is conserved in all such diagrams, again leading to a stable
proton. Although baryon and lepton number are both violated in the second
example, the proton does not decay rapidly. We emphasize that this reasoning
does not depend on whether one considers a specific model with fixed $U(1)_Y
\times U(1)'$ charges  or a scenario which imposes only a discrete subgroup
$\mathds Z^Y_3 \times \mathds Z_3$. 

To find $\mathcal Q$, we have to solve a homogeneous set of $J$ linear
equations, each equation corresponding to one allowed operator. Denoting the
baryon number of operator $j$ by $\mathcal B[j]$ and likewise for the 
lepton and the exotic quark number, we are looking for coefficients
$(b,\ell,k_1,k_2,k_3)$ which satisfy 
\beq
b \, \mathcal B[j] + \ell \, \mathcal L[j] + k_1  \, \mathcal K_1[j]+ k_2
 \, \mathcal K_2[j]+ k_3  \, \mathcal K_3[j] ~=~0 \ , \label{findingQ}
\eeq
for all $1 \leq j \leq J$. Having $J$ equations, at most five of them can
be linearly independent. The number of linearly independent equations is
called the rank $\mathfrak r$ of the set of equations. In the case where
$\mathfrak r = 5$, the only solution to eq.~(\ref{findingQ}) is
$b=\ell=k_1=k_2=k_3=0$, thus no conserved quantity~$\mathcal Q$ exists. If,
however, $\mathfrak r <5$, a non-trivial solution exists and with it a
conserved quantity $\mathcal Q$ is guaranteed. 

In the following section we fix the $\mathds Z_3$ symmetry among the MSSM
particles and scan over all 270 possible extensions of $\mathds Z_3^Y \times
\mathds Z_3$ which include the exotic quarks. For each case, we determine the
allowed operators and calculate the rank $\mathfrak r$, keeping only those
cases with $\mathfrak r<5$.

%%%%%%%%%%%%%%%%%%%%%%%%%%%%%%%%%%%%%%%%%%%%%%%%%%%%%%%%%%%
\section{Good $\bsym{\mathds Z_3^Y \times \mathds Z_3}$ Extensions}
\label{good}
%%%%%%%%%%%%%%%%%%%%%%%%%%%%%%%%%%%%%%%%%%%%%%%%%%%%%%%%%%%
Here we do not need to consider the case of $B_3$ since it already 
guarantees absolute proton stability as discussed in Section \ref{absolute}.
We will therefore only consider extensions of $M_3$ and $L_3$ in this section.

\vspace{5mm}
\noindent
{\bf $\bsym{M_3}$ Extensions}\\
Out of the 270 possible discrete charge assignments, only 20 lead to a reduced
rank. Interestingly, it is the discrete hypercharge $\mathds Z_3^Y$ that is
responsible for the occurrence of a non-trivial conserved quantity $\mathcal
Q$. This arises if 
\beq
q^Y[K_1]~=~q^Y[K_2]~=~q^Y[K_3]~=~q^Y[K]~=~0~\mbox{or}~2 \ ,\label{miracle}
\eeq
regardless of the charges $q[K_i]$ under $\mathds Z_3$. In these scenarios,
$\mathcal Q = \mathcal B$, so baryon number is conserved up to dimension five
operators including the exotic quarks. Therefore, the proton is sufficiently
stable.

It is instructive to figure out the reason for this peculiar
result. Under $M_3$, operators composed of only MSSM particles and right-handed
neutrinos do not violate baryon number (see Table~\ref{table2}). We will show
that the inclusion of exotic quarks does not allow the construction of
$\mathcal B$ violating operators up to dimension five which at the same time
conserve the generation independent discrete hypercharge $\mathds Z_3^Y$
displayed in Table~\ref{table4}. 
\TABLE[t]{
$\begin{array}{|c|c|c|c|c|c|c|c|c|c|}\hline
q^Y[Q] & q^Y[U^c] & q^Y[D^c] & q^Y[L] & q^Y[N^c] & q^Y[E^c] & q^Y[H_1] &
q^Y[H_2] & q^Y[K_i] & \phantom{\Big|}q^Y[K^c_i] \\ \hline\hline
1 & -1 & -1 & 0  & 0& 0& 0& 0 & q^Y[K] & -q^Y[K]\phantom{\Big|} \\\hline 
\end{array}$
\caption{\label{table4}The generation independent $\mathds Z_3^Y$
charges.}}
Invariance under $\mathds Z_3^Y$ requires
\beq
n_{Q} - n_{U^c} - n_{D^c} + q^Y[K] (n_{K} -  n_{K^c}) ~=~0~\mbox{mod}~3\ .\label{hyper}
\eeq
On the other hand, $SU(3)_C$ invariance demands
\beq
n_{Q} - n_{U^c} - n_{D^c} + n_{K} -  n_{K^c} ~=~0~\mbox{mod}~3\ .\label{color}
\eeq
Subtracting eq.~(\ref{color}) from eq.~(\ref{hyper}) yields
\beq
 (q^Y[K]-1) (n_{K} -  n_{K^c}) ~=~0~\mbox{mod}~3 \ ,\label{cond1}
\eeq
which for $q^Y[K]=0~\mbox{or}~2$ can only be satisfied if 
\beq
(n_{K} -  n_{K^c}) =
0~\mbox{mod}~3 \ .\label{cond1b}
\eeq
Then, eq.~(\ref{color}) simplifies to
\beq
n_{Q} - n_{U^c} - n_{D^c} ~=~0~\mbox{mod}~3\ ,\label{cond2}
\eeq
showing that baryon number violation requires at least three baryonic
fields. Without exotic particles, the symmetry $M_3$ ensures that no baryon
number violation occurs up to dimension five. Allowing for the
presence of exotic quarks in such an operator, we would need at least two of
them because of eq.~(\ref{cond1b}). An operator including three baryons (in
order to have baryon number violation) and two exotic quarks is, however,
suppressed by at least two powers of $M$. Hence, it is the
invariance under $SU(3)_C$ and $\mathds Z_3^Y$ that is responsible for baryon
number conservation in $M_3$ extensions with $q^Y[K]=0~\mbox{or}~2$.

It turns out that the same holds true for all $\mathds Z_N$ symmetries that 
have $\mathcal B$ conservation up to dimension five among the MSSM particles 
and the right-handed neutrinos, for instance the $\mathds Z_6$ symmetries 
$R_6^{}L_6^2~ (\cong M_3 \times R_2)$, and $R_6^3L_6^2~ (\cong L_3 \times R_2)$ 
\cite{Dreiner:2005rd,Luhn:2007gq} (see also Appendix~\ref{app-z6}).   

\vspace{5mm}
\noindent 
{\bf $\bsym{L_3}$ Extensions}\\
Scanning over the $\mathds Z_3^Y \times \mathds Z_3$ extensions of $L_3$ shows
that $\mathfrak r =5$ always. Therefore, with only the discrete symmetry at
our disposal, we do not obtain a conserved quantity $\mathcal Q$. However,
since $L_3$, to some extent, suggests the conservation of lepton number, one
could remove the $L_3$ invariant but lepton number violating operators
$N^cN^cN^c$ and $SN^cN^cN^c$ from the set of allowed operators (see Table \ref{table2}), and determine
$\mathfrak r$ for the remaining sets. The idea behind this procedure is that
one can easily forbid these two interactions with the underlying $U(1)'$ by
demanding $z[N^c]\neq 0$ and $z[N^c]\neq -z[S]/3$. Disregarding these
operators, we find that the rank is reduced to four in 55 of the 270 possible
extensions. The conserved quantity among the remaining operators is always
$\widetilde{\mathcal Q} = \mathcal L$. Not all of these discrete charge
assignments are compatible with the $[U(1)_Y]^2-U(1)'$ anomaly condition
$A_{111'}$, i.e. eq.~(42) in Ref.~\cite{Lee:2007fw}. With $N_H=1$ and 
the normalization where $y[Q]=1$, $A_{111'}$ translates to  
\beq
\sum_{i=1}^3 y[K_i]^2 ~=~36 \ .\label{sumy2}
\eeq
The only integer solutions are $(0,0,\sigma_3\!\cdot\!6)$ and
$(\sigma_1 \!\cdot\! 2,\sigma_2 \!\cdot\! 4,\sigma_3 \!\cdot\! 4)$, with
$\sigma_i =\pm 1$, as well as permutations thereof. Translated to the discrete
hypercharges, these solutions correspond to $(0,0,0)$ and $(\sigma_1 \!\cdot\!
2,\sigma_2 \!\cdot\! 1,\sigma_3 \!\cdot\! 1)$. With the implicit convention
that $-2 = 1$ and $-1 = 2$, all viable discrete hypercharges are therefore of
the form 
\beq
\Big(q^Y[K_1],q^Y[K_2],q^Y[K_3]\Big)~=~(0,0,0) ~~ \mbox{or} ~~ (\sigma_1
\!\cdot\! 1,\sigma_2 \!\cdot\! 1,\sigma_3 \!\cdot\! 1)  \ .\label{disY2}
\eeq
Out of the 55 $\mathds Z_3^Y \times \mathds Z_3$ extensions of $L_3$ which
reduce the rank, only 17 cases comply with eq.~(\ref{disY2}). They are listed
in Table~\ref{table5}.
\TABLE[t]{
\begin{tabular}{|c||c|c|c|c|c|c|c|c|c|c|}  \hline
$\!\!\!\!\begin{array}{lr} ~ & \!\mathds Z_3 \\[-2mm] 
             \mathds Z_3^Y\! &      ~  \end{array}\!\!\!\!$    & 
$\!\!(0,0,0)\!\!$ & $\!\!(1,1,1)\!\!$ & $\!\!(2,2,2)\!\!$ &
$\!\!(0,0,1)\!\!$ & $\!\!(0,0,2)\!\!$ &  $\!\!(0,1,1)\!\!$ &
$\!\!(1,1,2)\!\!$ & $\!\!(0,2,2)\!\!$ &   $\!\!(1,2,2)\!\!$ &
$\!\!(0,1,2)\!\!$ \\ \hline\hline

$\!\!(0,0,0)\!\!$ & $\checkmark~\smiley$ & $\checkmark~\smiley$ & $\checkmark~\smiley$ & & & & & & & \\ \hline

$\!\!(1,1,1)\!\!$ & $\checkmark$ & & & & & & & & & \\ \hline

$\!\!(2,2,2)\!\!$ & $\checkmark~\smiley$ & $\checkmark~\smiley$ & $\checkmark~\smiley$ & & & & & & & \\ \hline

$\!\!(1,1,2)\!\!$ & $\checkmark$ & & & $\checkmark$ & $\checkmark$ & & & & & \\ \hline

$\!\!(1,2,1)\!\!$ & $\checkmark$ & & & & & & & & & \\ \hline

$\!\!(2,1,1)\!\!$ & $\checkmark$ & & & & & & & & & \\ \hline

$\!\!(1,2,2)\!\!$ & $\checkmark$ & & & & & $\checkmark$ & & $\checkmark$ & & \\ \hline

$\!\!(2,1,2)\!\!$ & $\checkmark$ & & & & & & & & & \\ \hline

$\!\!(2,2,1)\!\!$ & $\checkmark$ & & & & & & & & & \\ \hline
\end{tabular}\caption{\label{table5}The $\mathds Z_3^Y \times \mathds Z_3$
  extensions of $L_3$. The discrete hypercharges $q^Y[K_i]$ that satisfy 
  eq.~(\ref{disY2}) are shown in the rows; the discrete charges $q[K_i]$ of 
  Table~\ref{table3} are given in the columns. The symbol $\checkmark$
  indicates that lepton number is violated only in the 
  operators $N^cN^cN^c$ and $SN^cN^cN^c$. A smiley $\smiley$ denotes cases
  where baryon number is only violated in $U^cD^cD^c$, $SU^cD^cD^c$, $QQQH_1$
  and $QQ{D^c}^\dagger$.}
}
The symbol $\checkmark$ indicates the 17 cases in which $\widetilde{\mathcal
Q} = \mathcal L$, i.e. those cases in which lepton number can only be violated
by $N^cN^cN^c$ and $SN^cN^cN^c$. For those 6 symmetries additionally marked
with the symbol $\smiley$, the only baryon number violating operators up to
dimension five are $U^cD^cD^c$, $SU^cD^cD^c$, $QQQH_1$ and $QQ{D^c}^\dagger$,
neither of which involves exotic fields. 
\begin{center}
$
\begin{array}{cll}
\checkmark :~ & \mathcal L~\mbox{violation~only~in} & N^cN^cN^c,~SN^cN^cN^c\ .
\\%[3mm] 
\smiley :~ &    \mathcal B~\mbox{violation~only~in} &
U^cD^cD^c,~SU^cD^cD^c,~QQQH_1,~QQ{D^c}^\dagger \ . 
\end{array}
$
\end{center}
The remaining $17-6=11$ cases violate baryon number also in many interactions 
involving exotic quarks.

For the symmetries in Table~\ref{table5} indicated by $\checkmark$, 
the proton can be stabilized by forbidding the two lepton number violating 
operators by the continuous
$U(1)'$. In the cases marked with the symbol $\smiley$, one could
alternatively control the four baryon number violating interactions; if, for
instance, $N^cN^cN^c$ is absent but $SN^cN^cN^c$ is allowed, one just has to
forbid the renormalizable term $U^cD^cD^c$ with the $U(1)'$ in order to make 
the proton sufficiently stable.

%%%%%%%%%%%%%%%%%%%%%%%%%%%%%%%%%%%%%%%%%%%%%%%%%%%%%%%%%%%
\section{$\bsym{L_3}$ Symmetric $\bsym{U(1)'}$ Models}
\label{l3symmetric}
%%%%%%%%%%%%%%%%%%%%%%%%%%%%%%%%%%%%%%%%%%%%%%%%%%%%%%%%%%%
In this section, we present phenomenologically viable and 
anomaly-free $SU(3)_C\times SU(2)_L\times U(1)_Y \times U(1)'$ models 
which have a $\mathds Z_3^Y \times \mathds Z_3$ extension of $L_3$ as 
a subgroup. We choose $N_H=1$, $a=1$ and $m=0$ or $m=-1$\footnote{Recall that 
up to now we have been assuming $n,m\ge 0$, so that any $1/M$ suppression
in the dimensionless couplings is coming solely from $\frac{S}{M}$ factors.
However, it can be readily seen from eqs.~(\ref{eqmu}) and (\ref{ABsup})
that negative values of $n$ and $m$ are also possible, and could be 
interpreted as a corresponding suppression due to $\frac{H_2H_1}{M^2}$ factors instead.}.
Additionally, we showcase two anomaly-free models which are incompatible 
with either proton longevity or the measured quark masses. 

Requiring $U(1)_Y \supset \mathds Z_3^Y$ entails integer hypercharges for all
particles, including the exotic quarks. Therefore eq.~(\ref{sumy2}) has only
a finite number of solutions. In our search for concrete models, we choose the
following 8+2 assignments 
\beq
(y[K_1],y[K_2],y[K_3]) ~=~ \left\{ \begin{array}{l} 
(\sigma_1 \!\cdot\! 2,\sigma_2\!\cdot\! 4,\sigma_3\!\cdot\! 4) \ , \\
(0,0,\sigma_3 \!\cdot\!  6) \ ,
\end{array} \right.
\eeq
with $\sigma_i=\pm 1$. All other possibilities are obtained from these by
relabeling the generations of the exotic quarks.

With regard to the $U(1)'$ symmetry, eq.~(\ref{BVmaster}) shows that one
particular charge assignment is accompanied by a two-dimensional space of
solutions which allow and forbid exactly the same operators\footnote{Note,
however, that two models with different $U(1)'$ charge assignments have 
different couplings to the $Z'$ and $\tilde{Z'}$.}. One dimension is spanned by
adding a certain amount of the hypercharge vector to the original charge
assignment, the other arises due to 
the choice of the overall normalization. To be explicit, we keep only those
assignments with $z[Q]=0$ and $|z[S]|=3$; the overall sign is fixed by
demanding compatibility of eq.~(\ref{defq}) with Table~\ref{table1}. In other
words, we take $\alpha + 3(1+m)\gamma = 0$ and $3\gamma = -1$ in eq.~(\ref{BVmaster}). In order
to end up with a $\mathds Z_3$ symmetry after $U(1)'$ breaking, the
charges $z[K_i]$ must be integers. As the cubic anomaly $[U(1)']^3$ is
quadratic in $z[K_i]$, we need to scan only over a finite number
of assignments $(z[K_1],z[K_2],z[K_3])$ to find all anomaly-free models. 
The phenomenologically viable models are listed in Table~\ref{table6}. 
\TABLE[t]{
$
\begin{array}{|c||r|r|r|r|r|r|r|r|r|r|r|r|r|r|}\hline
\rule[-5pt]{0pt}{18pt} &  \multicolumn{5}{|c|}{L_3\mbox{~models~with~} m=0} &
\multicolumn{9}{|c|}{L_3\mbox{~models~with~} m=-1}
\\\hline\hline
z[Q] & \multicolumn{5}{|c|}{\phantom{-}0~~} & \multicolumn{9}{|c|}{\phantom{-}0~~}  \rule[-5pt]{0pt}{18pt}\\%\hline
z[U^c] & \multicolumn{5}{|c|}{-6~~} & \multicolumn{9}{|c|}{-3~~}  \rule[-5pt]{0pt}{18pt}\\%\hline
z[D^c] & \multicolumn{5}{|c|}{\phantom{-}3~~} & \multicolumn{9}{|c|}{\phantom{-}0~~}  \rule[-5pt]{0pt}{18pt}\\%\hline
z[L] & \multicolumn{5}{|c|}{-1~~} & \multicolumn{9}{|c|}{-1~~}  \rule[-5pt]{0pt}{18pt}\\%\hline
z[N^c] & \multicolumn{5}{|c|}{-2~~} & \multicolumn{9}{|c|}{\phantom{-}1~~}  \rule[-5pt]{0pt}{18pt}\\%\hline
z[E^c] & \multicolumn{5}{|c|}{\phantom{-}4~~} & \multicolumn{9}{|c|}{\phantom{-}1~~}  \rule[-5pt]{0pt}{18pt}\\%\hline
z[H_1] & \multicolumn{5}{|c|}{-3~~} & \multicolumn{9}{|c|}{\phantom{-}0~~}  \rule[-5pt]{0pt}{18pt}\\%\hline
z[H_2] & \multicolumn{5}{|c|}{\phantom{-}6~~} & \multicolumn{9}{|c|}{\phantom{-}3~~}  \rule[-5pt]{0pt}{18pt}\\%\hline
z[S] & \multicolumn{5}{|c|}{-3~~} & \multicolumn{9}{|c|}{-3~~}  \rule[-5pt]{0pt}{18pt}\\\hline
& \mbox{I} & \mbox{~~II} & \mbox{~III} & \mbox{~~IV} & \mbox{V} & \mbox{VI} &
\mbox{VII} & \mbox{VIII} & \mbox{~IX} & \mbox{~~X} & \mbox{~XI} & \mbox{XII} &
\mbox{XIII} & \mbox{XIV}  \rule[-5pt]{0pt}{18pt}\\\hline
%
%%%%%\mbox{in Ref.}~{\cite{Lee:2007fw}} & \mbox{BV-I} & \mbox{} & \mbox{} & \mbox{} & \mbox{} & \mbox{BV-IV} &
\mbox{in Ref.}~{[16]} & \mbox{BV-I} & \mbox{} & \mbox{} & \mbox{} & \mbox{} & \mbox{BV-IV} &
\mbox{} & \mbox{} & \mbox{} & \mbox{} & \mbox{} & \mbox{} &
\mbox{} & \mbox{}  \rule[-5pt]{0pt}{18pt}\\\hline
z[K_1] & 3 & 2 & 1 & 1 & 2 & 3 & 1 & 2 & 2 & 1 & 2 & 2 & 1 & 1  \rule[-5pt]{0pt}{18pt} \\%\hline
z[K_2] & -3 & 2 & 1 & 1 & 2 & 0 & 1 & 2 & 2 & 1 & 1 & 1 & 2 & 2  \rule[-5pt]{0pt}{18pt} \\%\hline
z[K_3] & -3 & 8 & -5 & 8 & -5 & 0 & 4 & -1 & 4 & -1 & 4 & -1 & 4 & -1   \rule[-5pt]{0pt}{18pt} \\\hline
y[K_1] & 2 & 0 & 0 & 0 & 0 & 2 & 0 & 0 & 0 & 0 & 0 & 0 & 0 & 0  \rule[-5pt]{0pt}{18pt} \\%\hline
y[K_2] & -4 & 0 & 0 & 0 & 0 & -4 & 0 & 0 & 0 & 0 & 0 & 0 & 0 & 0  \rule[-5pt]{0pt}{18pt} \\%\hline
y[K_3] & -4 & 6 & -6 & 6 & -6 & -4 & 6 & -6 & 6 & -6 & 6 & -6 & 6 & -6   \rule[-5pt]{0pt}{18pt} \\\hline\hline
q[K_1] & 0 & 2 & 1 & 1 & 2 & 0 & 1 & 2 & 2 & 1 & 2 & 2 & 1 & 1  \rule[-5pt]{0pt}{18pt} \\%\hline
q[K_2] & 0 & 2 & 1 & 1 & 2 & 0 & 1 & 2 & 2 & 1 & 1 & 1 & 2 & 2  \rule[-5pt]{0pt}{18pt} \\%\hline
q[K_3] & 0 & 2 & 1 & 2 & 1 & 0 & 1 & 2 & 1 & 2 & 1 & 2 & 1 & 2   \rule[-5pt]{0pt}{18pt} \\\hline
q^Y[K_1] & 2 & 0 & 0 & 0 & 0 & 2 & 0 & 0 & 0 & 0 & 0 & 0 & 0 & 0  \rule[-5pt]{0pt}{18pt} \\%\hline
q^Y[K_2] & 2 & 0 & 0 & 0 & 0 & 2 & 0 & 0 & 0 & 0 & 0 & 0 & 0 & 0  \rule[-5pt]{0pt}{18pt} \\%\hline
q^Y[K_3] & 2 & 0 & 0 & 0 & 0 & 2 & 0 & 0 & 0 & 0 & 0 & 0 & 0 & 0   \rule[-5pt]{0pt}{18pt} \\\hline
\rule[-5pt]{0pt}{18pt}\mbox{class}&\multicolumn{3}{|c|}{\checkmark ~~~\smiley}&
\multicolumn{2}{|c|}{-} &  \multicolumn{3}{|c|}{\checkmark ~~~ \smiley} &
\multicolumn{6}{|c|}{-} \\\hline  
\end{array}
$
\caption{\label{table6}Phenomenologically viable $L_3$ models
with a $\mathds Z_3^Y \times \mathds Z_3$ subgroup. Six of them, indicated by
the symbols $\checkmark$ and $\smiley$, fall into a class where already the
discrete $\mathds Z_3^Y \times \mathds Z_3$ symmetry drastically limits the allowed
$\mathcal B$ and $\mathcal L$ violating operators, see
Table~\ref{table5}.} }
Comparing with Table~\ref{table5} shows that models I--III and VI--VIII belong
to the class ``$\checkmark~\smiley$'' with a $\mathds Z_3^Y \times \mathds
Z_3$ symmetry that allows only a few baryon and lepton number operators. Their
absence or presence in the specific model can be checked immediately. We find:
\begin{itemize}
\item {\bf I--III:} Neither $N^cN^cN^c$ nor $SN^cN^cN^c$ is allowed by the
  $U(1)'$, so lepton number is conserved up to dimension five. The only baryon
  number violating operator is $U^cD^cD^c$. Therefore, the proton is safe in
  these models. Up to a hypercharge shift and an overall minus sign, model~I
  is identical to the ``BV--I'' case in Ref.~\cite{Lee:2007fw}.
\item {\bf VI--VIII:} Here, lepton and baryon number are separately violated, but
  only in non-renormalizable operators, namely $SN^cN^cN^c$,
  $QQQH_1$ and $QQ{D^c}^\dagger$. Hence, any diagram that makes the proton
  decay is necessarily suppressed by at least two powers of
  $M$, leading to a sufficiently long  proton lifetime. Model~VI
  is equivalent to the ``BV--IV'' case in Ref.~\cite{Lee:2007fw}.
\end{itemize}
In order to see that the proton does not decay rapidly in the remaining
$14-6=8$ cases of Table~\ref{table6}, we need to construct all allowed
operators up to dimension five and determine the conserved quantity
$\mathcal Q$ for each model individually. We obtain:
\begin{itemize}
\item {\bf IV:} $\mathcal Q=\mathcal L - \mathcal K_3$. Baryon number is
  therefore violated (through $U^cD^cD^c$), but lepton number is
  conserved in processes where there is no external exotic quark. Thus 
  the proton is sufficiently stable.
\item {\bf V:} $\mathcal Q=\mathcal L + \mathcal K_3$. Same as for model IV.
\item {\bf IX--XIV:} $\mathcal Q=\mathcal K_3$. In these cases, the existence
of a conserved quantity $\mathcal Q$ does not guarantee a stable proton. We
must resort to the full list of baryon and lepton number violating operators
up to dimension five. It shows that, in all models, the only baryon number
violating operators are $QQQH_1$ and $QQ{D^c}^\dagger$. Lepton number, on the
other hand, is violated through $SN^cN^cN^c$ in all models, and additionally
through 
\beqn
&N^cK_1^\dagger K_2 \,,~N^cK^c_1{K^c_2}^\dagger\,,~SN^cK^c_1K_2~
&\quad \mbox{in~XI~and~XII}\ ,\nonumber \\
&N^c{K^c_1}^\dagger K^c_2 \,,~N^cK_1K_2^\dagger\,,~SN^cK_1K^c_2
&\quad \mbox{in~XIII~and~XIV}\ .\nonumber
\eeqn
Since baryon and lepton number are separately violated only at the
non-renormalizable level, the proton is safe in these models.
\end{itemize}

Having presented phenomenologically viable $\mathds Z_3^Y \times \mathds
Z_3$ models which are symmetric under $L_3$, we now discuss the shortcomings
of two anomaly-free $L_3$ symmetric $U(1)'$ charge assignments which lead to
contradictions with observations. Except for the exotic quarks, the $U(1)'$
charges in both cases are identical to those of models~I--V. The first case
has
$$
\Big(z[K_1],z[K_2],z[K_3] \Big) = (1,2,8) \ , \qquad
\Big(y[K_1],y[K_2],y[K_3] \Big) = (0,0,6) \ ,
$$
leading to no conserved quantity $\mathcal Q$. Up to dimension five,
$U^cD^cD^c$ is the only baryon number violating operator; lepton number is
violated in 
$$
E^cK_1K^c_3\,,~N^cK_1K^c_2\,,~N^cN^cK^c_1K_2 \ .
$$ From the latter two terms one can obtain the effective operator
$N^cN^cN^c$ at the loop level by contracting $K_i$ with $K^c_i$,
$i=1,2$. Therefore the diagram leading to proton decay is suppressed by only
one power of $M$ in this case.

In the second example, the exotic quarks have charges
$$
\Big(z[K_1],z[K_2],z[K_3] \Big) = (3,6,6) \ , \qquad
\Big(y[K_1],y[K_2],y[K_3] \Big) = (2,4,4) \ .
$$ 
This choice results in the conserved quantity $\mathcal Q = \alpha \mathcal L
+ \beta \mathcal K_1$. As lepton number is conserved, one might consider this
a physically acceptable charge assignment. However, the exotic quarks $K_{2}$
and $K_{3}$ mix with the up-type quarks through the superpotential operators
(for the sake of clarity we suppress all generational indices)
\beq
M \, KU^c \ , \qquad \frac{1}{M} \, S H_2 Q K^c \
. \label{mix}
\eeq
After $S$ and $H_2$ acquire their vevs, we obtain the mass terms
\beq
\begin{pmatrix}
U & K
\end{pmatrix}
\cdot
\begin{pmatrix} 
c_{11}\, \langle  H_2 \rangle 
&c_{12}\,\frac{\langle S\rangle \langle H_2 \rangle}{M} \\[2mm]
c_{21}\, M & c_{22}\, \langle S \rangle
\end{pmatrix}
\cdot
\begin{pmatrix}
U^c \\ K^c
\end{pmatrix} ,
\eeq
with eigenvalues of the order
\beq
c_{21}\, M \ , \qquad 
\left(\frac{c_{11}c_{22}}{c_{21}} - c_{12}\right) \cdot
\frac{\langle S\rangle \langle H_2 \rangle}{M}  \ .
\eeq
Assuming no artificially small value for the coupling coefficient $c_{21}$,
the second mass eigenvalue is way too small to account for the up-type quark masses. 
Therefore,
a scenario in which the exotic quarks mix with the observed ones as in
eq.~(\ref{mix}) would be highly unnatural.

%%%%%%%%%%%%%%%%%%%%%%%%%%%%%%%%%%%%%%%%%%%%%%%%%%%%%%%%%%%
\section{Models with $\bsym{\mathds Z_3^Y \times \mathds Z_{N>3}}$}
\label{z3yzn}
%%%%%%%%%%%%%%%%%%%%%%%%%%%%%%%%%%%%%%%%%%%%%%%%%%%%%%%%%%%
We can now relax
the requirement of integer $U(1)'$ charges  for the exotic quarks. After
rescaling the charges, this is tantamount to looking for scenarios where
$U(1)' \rightarrow \mathds Z_{N>3}$. Indeed, we find many such anomaly-free
models. For $B_3$ and $L_3$, some are given in Ref.~\cite{Lee:2007fw}. It is
the purpose of this section to argue for the stability of the proton in the
models of Ref.~\cite{Lee:2007fw}, as well as in some new models featuring the discrete
symmetry~$M_3$. Concerning the $B_3$ models of Ref.~\cite{Lee:2007fw}, we have
already  shown in Section~\ref{absolute} that the proton is absolutely
stable. The $U(1)'$ charge assignments which we are going to discuss here are
only the $L_3$ and $M_3$ cases given in Table~\ref{table7}. 
The primed models are related to the unprimed
ones by simultaneously changing $y[K_i] \leftrightarrow y[K^c_i]$ and
$z[K_i] \leftrightarrow z[K^c_i]$; the thus obtained charge assignments are
also anomaly-free because the anomaly coefficients do not distinguish between
$SU(3)_C$ triplets and antitriplets (see Ref.~\cite{Lee:2007fw}).

In order to determine whether a model is consistent with the longevity of the
proton, we take the same approach as in the previous section. First we
filter out the information contained in the $\mathds Z_3^Y \times \mathds Z_N$
subgroup. If the rank $\mathfrak r$ of the set of homogeneous linear equations
derived from the allowed operators, see eq.~(\ref{findingQ}), is less than 5,
we have to find the conserved quantity $\mathcal Q$ for these scenarios. In
some cases, no further effort has to be made because the discrete symmetry
already stabilizes the proton. However, often we have to take a second step
and determine the conserved quantity $\mathcal Q$ of the specific model
(i.e. using the exact $U(1)'$ charges). If that also fails, 
we need to investigate explicitly all baryon and lepton number
violating operators up to dimension five. 
\TABLE[t]{
{$
\begin{array}{|c||r|r|r|r|r|r|r|r|r|r|r|r|r|} \hline
\rule[-5pt]{0pt}{18pt}& \multicolumn{8}{|c|}{L_3\mbox{~models}} 
& \multicolumn{5}{|c|}{M_3\mbox{~models}} \\ \hline\hline
\rule[-5pt]{0pt}{18pt}z[Q]&\multicolumn{2}{|r|}{ 0 ~\quad}& \multicolumn{2}{|r|}{ 0 ~~\quad} &\multicolumn{4}{|r|}{ 0  ~\qquad\qquad} &

\multicolumn{1}{|c}{~}&\multicolumn{1}{r}{ 0~ }&\multicolumn{1}{r|}{~}&\multicolumn{2}{|r|}{ 0 ~~~\quad} 
 \\
\rule[-5pt]{0pt}{18pt}z[U^c]&\multicolumn{2}{|r|}{ -18 ~\quad}& \multicolumn{2}{|r|}{ -90 ~~\quad}&\multicolumn{4}{|r|}{-9 ~\qquad\qquad} & 

\multicolumn{1}{|c}{~}&\multicolumn{1}{r}{6~ }&\multicolumn{1}{r|}{~}&\multicolumn{2}{|r|}{15 ~~~\quad} 
 \\
\rule[-5pt]{0pt}{18pt}z[D^c]&\multicolumn{2}{|r|}{ 9  ~\quad}&  \multicolumn{2}{|r|}{ 45  ~~\quad}   &\multicolumn{4}{|r|}{  0  ~\qquad\qquad} & 

\multicolumn{1}{|c}{~}&\multicolumn{1}{r}{ 3~ }&\multicolumn{1}{r|}{~}&\multicolumn{2}{|r|}{-6 ~~~\quad} 
 \\
\rule[-5pt]{0pt}{18pt}z[L]&\multicolumn{2}{|r|}{ -3  ~\quad}& \multicolumn{2}{|r|}{ -15  ~~\quad}&\multicolumn{4}{|r|}{ -3 ~\qquad\qquad} & 

\multicolumn{1}{|c}{~}&\multicolumn{1}{r}{  3~}&\multicolumn{1}{r|}{~}&\multicolumn{2}{|r|}{ 3 ~~~\quad} 
 \\
\rule[-5pt]{0pt}{18pt}z[N^c]&\multicolumn{2}{|r|}{-6 ~\quad}& \multicolumn{2}{|r|}{-30 ~~\quad}  &\multicolumn{4}{|r|}{ 3  ~ \qquad\qquad} & 

\multicolumn{1}{|c}{~}&\multicolumn{1}{r}{-6~ }&\multicolumn{1}{r|}{~}&\multicolumn{2}{|r|}{ 3~~~\quad} 
 \\
\rule[-5pt]{0pt}{18pt}z[E^c]&\multicolumn{2}{|r|}{12 ~\quad}& \multicolumn{2}{|r|}{60 ~~\quad}  &\multicolumn{4}{|r|}{ 3   ~\qquad\qquad} & 

\multicolumn{1}{|c}{~}&\multicolumn{1}{r}{ 0~ }&\multicolumn{1}{r|}{~}&\multicolumn{2}{|r|}{-9 ~~~\quad} 
 \\
\rule[-5pt]{0pt}{18pt}z[H_1]&\multicolumn{2}{|r|}{-9 ~\quad}& \multicolumn{2}{|r|}{-45 ~~\quad} &\multicolumn{4}{|r|}{ 0   ~\qquad\qquad} & 

\multicolumn{1}{|c}{~}&\multicolumn{1}{r}{-3~  }&\multicolumn{1}{r|}{~}&\multicolumn{2}{|r|}{6~~~\quad} 
 \\
\rule[-5pt]{0pt}{18pt}z[H_2]&\multicolumn{2}{|r|}{ 18  ~\quad}& \multicolumn{2}{|r|}{ 90  ~~\quad}   &\multicolumn{4}{|r|}{9  ~\qquad\qquad} & 

\multicolumn{1}{|c}{~}&\multicolumn{1}{r}{-6~ }&\multicolumn{1}{r|}{~}&\multicolumn{2}{|r|}{-15~~~\quad} 
 \\
\rule[-5pt]{0pt}{18pt}z[S]&\multicolumn{2}{|r|}{-9 ~\quad}& \multicolumn{2}{|r|}{-45 ~~\quad} &\multicolumn{4}{|r|}{ -9  ~\qquad\qquad} & 

\multicolumn{1}{|c}{~}&\multicolumn{1}{r}{ 9~ }&\multicolumn{1}{r|}{~}&\multicolumn{2}{|r|}{ 9 ~~~\quad} 
 \\\hline
\rule[-5pt]{0pt}{18pt}& \mbox{(i)} & \mbox{(i')} & \mbox{(ii)} & \mbox{(ii')}& \mbox{(iii)}& \mbox{(iii')}& \mbox{(iv)} & \mbox{(iv')}& \mbox{(v)} & \mbox{(vi)} & \mbox{(vii)} & \mbox{(viii)}& \mbox{(ix)}  \\\hline 
%
%%%\rule[-5pt]{0pt}{18pt} \mbox{in Ref.}~\cite{Lee:2007fw}& \mbox{BV-II} & \mbox{BV-II'} & \mbox{BV-III} & \mbox{BV-III'}& \mbox{BV-V}& \mbox{BV-V'}& \mbox{BV-VI} & \mbox{BV-VI'}& \mbox{} & \mbox{} & \mbox{} & \mbox{}& \mbox{}  \\\hline
\rule[-5pt]{0pt}{18pt} \mbox{in Ref.}~{[16]}& \mbox{BV-II} & \mbox{BV-II'} & \mbox{BV-III} & \mbox{BV-III'}& \mbox{BV-V}& \mbox{BV-V'}& \mbox{BV-VI} & \mbox{BV-VI'}& \mbox{} & \mbox{} & \mbox{} & \mbox{}& \mbox{}  \\\hline 
\rule[-5pt]{0pt}{18pt}z[K_1] & 13&-4&47&-2&7&2&5&4&-5&-5&-5&-11&-11 \\ 
\rule[-5pt]{0pt}{18pt}z[K_2] & -8&17&-40&85&1&8&-1&10&-5&-2&-2&7&-13 \\
\rule[-5pt]{0pt}{18pt}z[K_3] & -8&17&-49&94&-2&11&-1&10&-8&-2&-7&-13&-16 \\\hline
\rule[-5pt]{0pt}{18pt}y[K_1] & 2 & -2& 2&-2& 2&-2& 2&-2&0& 2&2&2&2 \\
\rule[-5pt]{0pt}{18pt}y[K_2] & -4 & 4 & -4 &4& -4 &4& -4 &4&0& -4& -4& -4& 4 \\
\rule[-5pt]{0pt}{18pt}y[K_3] & -4& 4 & -4 &4& -4 &4& -4 &4&6& -4& 4& 4& 4\\\hline\hline
\rule[-5pt]{0pt}{18pt}q^Y[K_1] & 2&1&2&1&2&1&2&1&0&2&2&2&2 \\
\rule[-5pt]{0pt}{18pt}q^Y[K_2] & 2&1&2&1&2&1&2&1&0&2&2&2&1 \\
\rule[-5pt]{0pt}{18pt}q^Y[K_2] & 2&1&2&1&2&1&2&1&0&2&1&1&1 \\\hline
\end{array}
$}
\caption{\label{table7}Some models with $U(1)' \rightarrow \mathds Z_{N>3}$.}
}

\vspace{5mm}
\noindent{\bf{$\bsym{L_3}$ Models}}\\
With only the discrete symmetry $\mathds Z_3^Y \times \mathds Z_N$ at hand,
the rank $\mathfrak r$ reduces only in cases (ii)/(ii') of
Table~\ref{table7}; we get the conserved quantity $\mathcal Q=\mathcal K_1 -
2\mathcal K_3$. Even if we disregard the operators $N^cN^cN^c$ and
$SN^cN^cN^c$, $\mathfrak r$ is not reduced in the other six cases. For
(ii)/(ii'), at the level of the discrete symmetry, we obtain 
$\widetilde{\mathcal Q}= \alpha (3 \mathcal L \mp \mathcal K_2) + \beta
(\mathcal K_1 - 2 \mathcal K_3)$, where the upper sign holds true for the
unprimed model and the lower for the primed one. We stick to this convention
throughout this section. Since the actual charges of
models (ii)/(ii') allow neither $N^cN^cN^c$ nor $SN^cN^cN^c$, lepton
number is conserved in all processes without external exotic quarks. So models
(ii)/(ii') are phenomenologically acceptable.

For the remaining six $L_3$ cases, we need to consider the exact $U(1)'$ charge
assignments. The obtained conserved quantities for the models are:

\begin{center}
$
\begin{array}{rlcl}
& \quad\mbox{including}~ (S)N^cN^cN^c 
& ~~~ & \quad\mbox{excluding}~ (S)N^cN^cN^c \\[2mm]
\mbox{(i)/(i'):}~~~ 
& \mathcal Q = 3\mathcal L \mp (\mathcal K_1 -2\mathcal K_2-2\mathcal K_3) \ ,
&&\widetilde{\mathcal Q}=3\mathcal L \mp(\mathcal K_1-2\mathcal K_2-2\mathcal K_3)\ , \\%[1mm]
%
%\mbox{(ii):}~~~ 
%& \mathcal Q = \mathcal K_1 + \mathcal K_2 + \mathcal K_3 \ ,
%&& \widetilde{\mathcal Q} = \mathcal K_1 + \mathcal K_2 + \mathcal K_3 \ ,
%\\%[1mm]
%
\mbox{(iii)/(iii'):}~~~
& \mathcal Q = \beta \mathcal K_1 + \gamma (\mathcal K_2 + \mathcal K_3)\ ,
&&  \widetilde{\mathcal Q} = \alpha (\mathcal L \pm \mathcal K_3) + \beta
\mathcal K_1 + \gamma (\mathcal K_2 + \mathcal K_3) \ ,
\\%[1mm]
\mbox{(iv)/(iv'):}~~~
&\mathcal Q = \mathcal K_1 - 2\mathcal K_2 - 2\mathcal K_3 \ ,
&& \widetilde{\mathcal Q} = \alpha \mathcal L + \beta (\mathcal K_1 -
2\mathcal K_2 - 2\mathcal K_3 ) \ .
\end{array}
$
\end{center}

In models (i)/(i'), the proton is safe due to $\mathcal L$ conservation in
processes with no external~$K_i$. The other models violate $\mathcal L$. 

However, for models (iv)/(iv') the only lepton number violation occurs in
$SN^cN^cN^c$. We must therefore determine the baryon number violating
operators. It is worth pointing out that already at the level of the discrete
$\mathds Z_3^Y \times Z_N$ symmetry, in all eight $L_3$ cases 
the only BV operators are precisely the $\smiley$ operators from Section~\ref{good}:
\beq
U^cD^cD^c \ , \quad SU^cD^cD^c \ , \quad QQQH_1\ , \quad QQ{D^c}^\dagger \ .
\eeq
The specific charges of models (iv)/(iv') forbid the first two terms, so both
$\mathcal L$ and $\mathcal B$ are only violated separately at the
non-renormalizable level. Hence, these models have a sufficiently stable
proton. 

Concerning models (iii)/(iii'), we must additionally determine all lepton
number violating operators up to dimension five:
\begin{center}
$
\begin{array}{rlcl}
& \mathcal B~\mbox{violation} &~~~&  \mathcal L~\mbox{violation} \\[2mm]
%\mbox{(ii):}~~~&U^cD^cD^c \ , 
%&& E^cK^c_1K_3\,,\,N^cK^c_2K_3\,,\,N^cN^cK_2K^c_3 \ , \\%[1mm]
\mbox{(iii):}~~~ & QQQH_1\,,\,QQ{D^c}^\dagger\ ,
&& N^cK^\dagger_2
K_3\,,\,N^cK^c_2{K^c_3}^\dagger\,,\,SN^cK^c_2K_3\,,\,SN^cN^cN^c\ , \\%[2mm]
\mbox{(iii'):}~~~ & QQQH_1\,,\,QQ{D^c}^\dagger\ ,
&& N^c{K^c_2}^\dagger K^c_3\,,\,N^cK_2K_3^\dagger\,,\,SN^cK_2K^c_3\,,\,SN^cN^cN^c\ .
\end{array}
$
\end{center}
We see that in models (iii)/(iii') baryon and lepton number are violated
separately only at the non-renormalizable level. So these are also viable
charge assignments. 

\vspace{5mm}
\noindent{\bf{$\bsym{M_3}$ Models}}\\
The symmetry $M_3$ forbids baryon number violation among the MSSM particles
and the right-handed neutrinos. For models (v) and (vi), the discrete
hypercharges satisfy eq.~(\ref{miracle}), which anticipates that $\mathcal B$
is also conserved at the level of the subgroup $\mathds Z_3^Y \times \mathds
Z_9$ once we add the exotic quarks. For the remaining
three cases (vii)$-$(ix), the discrete symmetry of the models also leads to the
conserved quantity $\mathcal Q=\mathcal B$. Therefore, all five $M_3$ models
given in Table~\ref{table7} have a stable proton.

%%%%%%%%%%%%%%%%%%%%%%%%%%%%%%%%%%%%%%%%%%%%%%%%%%%%%%%%%%%
\section{Summary and Conclusion}
\label{conclusion}
%%%%%%%%%%%%%%%%%%%%%%%%%%%%%%%%%%%%%%%%%%%%%%%%%%%%%%%%%%%
In this article, we investigated the issue of proton stability in the
general UMSSM with $R$-parity violation. The proton decay problem 
may arise due to two reasons. First, in the absence of $R$-parity, 
one might expect the usual RPV couplings to destabilize the proton. 
However, the LV-BV separation \cite{Lee:2007fw} ensures that
the dangerous LV and BV couplings cannot coexist, so that the
proton is safe from operators involving MSSM fields, even at the 
non-renormalizable level. The second, much more severe problem
arises due to the presence of light exotics, which are needed to
render the $U(1)'$ gauge symmetry free of anomalies. The exotics 
themselves may have LV and/or BV interactions, posing a serious problem 
for the stability of the proton. Nevertheless, we have identified 
several classes of models where the exotics are relatively harmless 
with respect to the proton decay issue. 

A central element in our analysis was the concept of discrete gauge symmetries.
After the spontaneous breaking of the $U(1)'$ gauge symmetry, 
any charge assignment automatically leads to a remnant $\mathds Z_N$ symmetry. 
Furthermore, there is an analogous $\mathds Z_N^Y$ discrete symmetry 
which is left over after the breaking of the hypercharge gauge group $U(1)_Y$.
We found that the knowledge of these discrete symmetries provides a powerful 
tool in arguing for the stability of the proton. Our main results are 
pictorially summarized in Fig.~\ref{roadmap}, where we present the main
steps one has to follow in deciding whether a particular UMSSM model is safe with 
respect to proton decay or not. We should stress that Fig.~\ref{roadmap} 
can be applied only to anomaly-free UMSSM models with a minimal exotic content,
i.e. three generations of $SU(2)_L$-singlet exotic quarks. Our method, however, 
can be easily generalized to the case of non-minimal exotic sectors as well.

%%%%%%%%%%%%%%%%%%%% FIGURE %%%%%%%%%%%%%%%%%%
\FIGURE{
\includegraphics[width=0.88\textheight,angle=90]{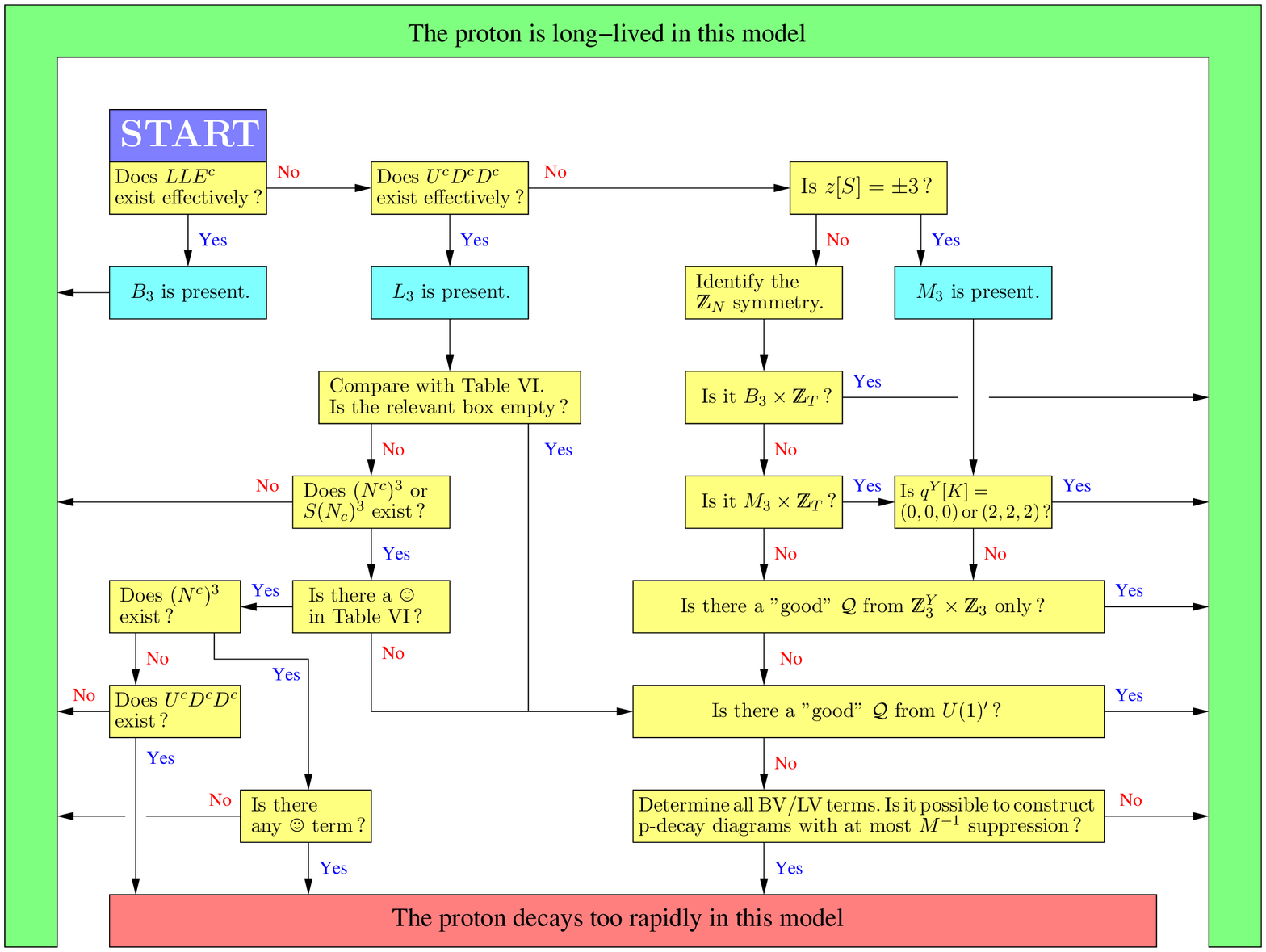}
\caption{A roadmap for RPV UMSSM model building.
\label{roadmap} } }
%%%%%%%%%%%%%%%%%%%%%%%%%%%%%%%%%%%%%%%%%%%%

In Section~\ref{absolute} we identified four symmetries 
($B_3$, $P_6$, $R^{}_{12}L^4_{12}$, $R^{5}_{12}L^8_{12}$) 
which render the proton absolutely stable. Fig.~\ref{roadmap}
confirms that the shortest path to the stable proton outcome 
is when the model exhibits a $B_3$ discrete symmetry.
For other discrete symmetries, knowing the couplings of the exotic quarks to
the MSSM particles and the right-handed neutrinos is essential. 
For this reason, we have extended the concept of a discrete 
symmetry to the exotic sector. Since the hypercharge of the new exotic
particles is also unknown, we introduced the notion of a discrete hypercharge. 
Scanning all possible $\mathds Z_3^Y \times \mathds Z_3$ extensions
of lepton triality~$L_3$ and matter triality~$M_3$, we found many cases in
which the discrete symmetry forbids (most of) the $SU(3)_C\times SU(2)_L$
invariant baryon and/or lepton number violating operators up to dimension
five. Their absence for a particular $U(1)'$ charge assignment ensures a
stable proton. This is confirmed by Fig.~\ref{roadmap}, which offers several 
alternative paths to the stable proton outcome, which rely primarily on the 
extended discrete symmetries encoded in the model.

Models which do not fall into these ``good'' categories need to be further
scrutinized. A method which we found very useful in classifying the 
remaining possibilities is the following. We generate all possible 
operators up to dimension five which are invariant under either
$SU(3)_C \times SU(2)_L \times U(1)_Y \times U(1)'$ or its discrete version 
$SU(3)_C \times SU(2)_L \times \mathds Z_3^Y \times \mathds Z_N $.
We then look for a conserved quantity $\mathcal Q$ among the set of those 
operators. As the proton cannot decay into exotic quarks, one can often 
argue for a sufficiently stable proton solely on the basis of $\mathcal Q$.
As evidenced from Fig.~\ref{roadmap} and some of our examples in 
Secs.~\ref{good}-\ref{z3yzn}, this can be often the case with 
$L_3$ and $M_3$ UMSSM models. Only for a few remaining model cases, 
it is necessary to explicitly write down all baryon and lepton number 
violating operators in order to verify whether the proton is stable.

Our results show that in spite of the presence of light exotics at the TeV scale,
the anomaly-free RPV UMSSM is a phenomenologically viable 
alternative to more conventional versions of low energy supersymmetry. 
It is instructive that a consistent model has {\em three} new elements 
in comparison to the (N)MSSM: 
(1) new $U(1)'$ gauge interactions and the associated
gauge particles and their superpartners; (2) RPV interactions and (3) 
new exotic isosinglet quarks and squarks at the TeV scale \cite{Kang:2007ib}.
One should therefore be on the lookout for such signatures
during the upcoming runs at the Large Hadron Collider at CERN.

\section*{Acknowledgments}
We thank Graham Ross for stimulating discussions.
HL and KM are supported by the Department of Energy under grant
DE-FG02-97ER41029. The work of CL is supported by the University of 
Florida through the Institute for Fundamental Theory.

\appendix

%%%%%%%%%%%%%%%%%%%%%%%%%%%%%%%%%%%%%%%%%%%%%%%%%%%%%%%%%%%
\section{Good $\bsym{\mathds Z_3^Y \times \mathds Z_6}$ Extensions}
\label{app-z6}
%%%%%%%%%%%%%%%%%%%%%%%%%%%%%%%%%%%%%%%%%%%%%%%%%%%%%%%%%%%
We have pointed out in Section~\ref{good} that those extensions of the
$\mathds Z_6$ symmetries $R_6^{}L_6^2~ (\cong M_3 \times R_2)$, and $R_6^3L_6^2~(\cong L_3 \times R_2)$
which have the discrete hypercharges of eq.~(\ref{miracle}) conserve baryon
number up to dimension five operators. The complete $\mathds Z_3^Y \times
\mathds Z_6$ scan reveals that no additional good symmetries that comply with
eq.~(\ref{disY2}) are obtained for $R_6^{}L_6^2$.

The situation changes dramatically when scanning over the extensions of the
symmetry $R_6^3L_6^2$. Out of the $27\times 56 = 1512$ possible discrete charge 
assignments\footnote{Concerning the $\mathds Z_6$
sector, there are 6 cases with identical $q[K_i]$, $6\times 5 = 30$ cases
where two $q[K_i]$ are identical, and finally $\frac{6\cdot 5\cdot 4}{3!}=20$
cases with all three discrete charges different from each other. This adds up
to 56 different $\mathds Z_6$ charge assignments for the exotic quarks.}, 
1298 have rank
$\mathfrak r$ smaller than 5, leading to a conserved quantity $\mathcal
Q$. Filtering out those cases which satisfy eq.~(\ref{disY2}), we are left
with 415 cases with non-trivial $\mathcal Q$. For illustration, we list those
16 scenarios in which the rank is reduced to $\mathfrak r =2$, together with
the corresponding conserved quantity  $\mathcal Q$, in Table~\ref{tableA1}.
$\alpha$, $\beta$, $\gamma$ are free real parameters. Therefore, one actually
has three independent conserved quantities in these scenarios.
\TABLE{
\begin{tabular}{|c|c||c|}\hline
$(q^Y[K_1],q^Y[K_2],q^Y[K_3])$ & $(q[K_1],q[K_2],q[K_3])$ 
& $\mathcal Q\phantom{\Big|}$ \\\hline\hline 

$\phantom{\Big|}\!(0,0,0)$\,or\,$(2,2,2)\phantom{\Big|}$ &
$(1,1,1)$\,or\,$(3,3,3)$\,or\,$(5,5,5)\!$ & 
$\alpha \mathcal B + \beta \mathcal L 
+ \gamma (\mathcal K_1 + \mathcal K_2 + \mathcal K_3)$  \\\hline 

$\phantom{\Big|}(1,1,2)\phantom{\Big|}$ &  $(3,3,3)$\,or\,$(3,3,5)$ &
$\alpha(\mathcal B + \mathcal K_1 + \mathcal K_2) + \beta \mathcal L  
+ \gamma \mathcal K_3$ \\\hline 

$\phantom{\Big|}(1,2,1)\phantom{\Big|}$ & $(3,3,3)$ &
$\alpha(\mathcal B + \mathcal K_1 + \mathcal K_3) + \beta \mathcal L  
+ \gamma \mathcal K_2 $ \\\hline 

$\phantom{\Big|}(2,1,1)\phantom{\Big|}$ & $(3,3,3)$\,or\,$(1,3,3)$ &
$\alpha(\mathcal B + \mathcal K_2 + \mathcal K_3) + \beta \mathcal L  
+ \gamma \mathcal K_1 $ \\\hline 

$\phantom{\Big|}(1,2,2)\phantom{\Big|}$ & $(3,3,3)$\,or\,$(3,5,5)$ &
$\alpha (\mathcal B + \mathcal K_1) + \beta \mathcal L  
+ \gamma (\mathcal K_2 + \mathcal K_3)$ \\\hline 

$\phantom{\Big|}(2,1,2)\phantom{\Big|}$ & $(3,3,3)$ &
$\alpha (\mathcal B + \mathcal K_2) + \beta \mathcal L  
+ \gamma (\mathcal K_1 + \mathcal K_3)$ \\\hline 

$\phantom{\Big|}(2,2,1)\phantom{\Big|}$ & $(3,3,3)$\,or\,$(1,1,3)$ &
$\alpha (\mathcal B + \mathcal K_3) + \beta \mathcal L  
+ \gamma (\mathcal K_1 + \mathcal K_2)$ \\\hline 
\end{tabular}\caption{\label{tableA1}$\mathds Z_3^Y \times \mathds Z_6$
extensions of $R_6^3L_6^2$ which lead to three conserved quantities $\mathcal
Q$.} 
}

%%%%%%%%%%%%%%%%%%%%%%%%%%%%%

\end{document}